\documentclass[conference]{IEEEtran}
\IEEEoverridecommandlockouts
\usepackage{cite}
\usepackage{amsmath,amssymb,amsfonts}
\usepackage{graphicx}
\usepackage{textcomp}
\usepackage{algpseudocode}
\usepackage{algorithm}
\usepackage{url}
\usepackage{xcolor}
\usepackage{multirow}
\usepackage{subfig}
\newtheorem{definition}{\textbf{Definition}} 
\newtheorem{insight}{\textbf{Insight}} 

\def\BibTeX{{\rm B\kern-.05em{\sc i\kern-.025em b}\kern-.08em
    T\kern-.1667em\lower.7ex\hbox{E}\kern-.125emX}}
\begin{document}

\title{
HaS: Accelerating RAG through Homology-Aware Speculative Retrieval
    \thanks{
    Corresponding Author: Weiwei Lin.
    
    This work was supported by the Shandong Provincial Natural Science Foundation (ZR2024LZH012), the Guangxi Key Research and Development Project (2024AB02018), the Major Key Project of PCL, China (PCL2025A11 and PCL2025A08), the New Generation Artificial Intelligence National Science and Technology Major Project (2025ZD0123605), the National Natural Science Foundation of China (62402198), and the Basic and Applied Basic Research Foundation of Guangzhou (2025A04J2212)
    }
    
}

\makeatletter
\newcommand{\linebreakand}{%
  \end{@IEEEauthorhalign}
  \hfill\mbox{}\par
  \mbox{}\hfill\begin{@IEEEauthorhalign}
}
\makeatother

\author{
\IEEEauthorblockN{1\textsuperscript{st} Peng Peng}
\IEEEauthorblockA{
    \textit{South China University of Technology}\\
    Guangzhou, China \\
}
\IEEEauthorblockA{
    \textit{Pengcheng Laboratory}\\
    Shenzhen, China \\
    pengp@pcl.ac.cn \\
}
\and
\IEEEauthorblockN{2\textsuperscript{nd} Weiwei Lin}
\IEEEauthorblockA{
    \textit{South China University of Technology}\\
    Guangzhou, China \\
}
\IEEEauthorblockA{
    \textit{Pengcheng Laboratory}\\
    Shenzhen, China \\
    linww@scut.edu.cn \\
}
\and
\IEEEauthorblockN{3\textsuperscript{rd} Wentai Wu}
\IEEEauthorblockA{
    \textit{Jinan University}\\
    Guangzhou, China \\
    wentaiwu@jnu.edu.cn \\
}
\linebreakand
\IEEEauthorblockN{4\textsuperscript{th} Xinyang Wang}
\IEEEauthorblockA{
    \textit{Beijing Forestry University} \\
    Beijing, China \\
    wxyyuppie@bjfu.edu.cn \\
}
\and
\IEEEauthorblockN{5\textsuperscript{th} Yongheng Liu}
\IEEEauthorblockA{
    \textit{Pengcheng Laboratory}\\
    Shenzhen, China \\
    yongheng.liu@pcl.ac.cn \\
}
}

\maketitle

\begin{abstract}
Retrieval-Augmented Generation (RAG) expands the knowledge boundary of large language models (LLMs) at inference by retrieving external documents as context. However, retrieval becomes increasingly time-consuming as the knowledge databases grow in size. Existing acceleration strategies either compromise accuracy through approximate retrieval, or achieve marginal gains by reusing results of strictly identical queries. We propose HaS, a homology-aware speculative retrieval framework that performs low-latency speculative retrieval over restricted scopes to obtain candidate documents, followed by validating whether they contain the required knowledge. The validation, grounded in the homology relation between queries, is formulated as a homologous query re-identification task: once a previously observed query is identified as a homologous re-encounter of the incoming query, the draft is deemed acceptable, allowing the system to bypass slow full-database retrieval. Benefiting from the prevalence of homologous queries under real-world popularity patterns, HaS achieves substantial efficiency gains. Extensive experiments demonstrate that HaS reduces retrieval latency by 23.74\% and 36.99\% across datasets with only a 1-2\% marginal accuracy drop. As a plug-and-play solution, HaS also significantly accelerates complex multi-hop queries in modern agentic RAG pipelines. Source code is available at: \url{https://github.com/ErrEqualsNil/HaS}.
\end{abstract}


\section{Introduction}
RAG well complements the insufficient knowledge coverage of LLMs \cite{rag}. Instead of relying on parametric memory, RAG retrieves factual documents from external databases to augment LLM responses through input prompts. Proved effective in reducing hallucinations, RAG has already been adopted widely in the industry for both general-purpose and domain-specific chat applications. 

\begin{figure}[!htbp]
  \centering
  \includegraphics[width=0.8\linewidth]{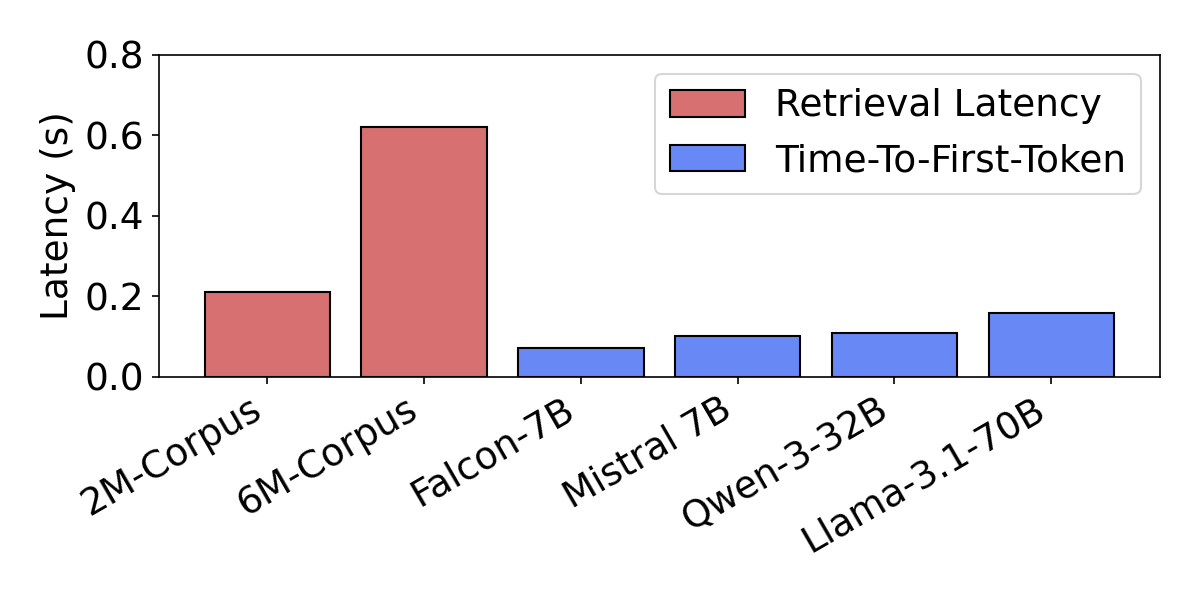}
  \caption{Retrieval is much slower than generation, as revealed by the comparison between retrieval latency and the time-to-first-token for a bare LLM.}
  \label{fig:ttft}
\end{figure}

Despite extensive efforts to reduce LLM prefilling latency caused by longer RAG prompts \cite{yao2024a, jin2024, cheng2024b}, recent studies have highlighted that the retrieval, implemented as an exact nearest neighbor search (ENNS), has emerged as a dominant latency bottleneck, significantly prolonging the time-to-first-token (TTFT) and degrading user experience \cite{quinn2025}. As global knowledge repositories continue to scale, this challenge becomes increasingly pronounced. As illustrated in Fig. \ref{fig:ttft}, our profiling reveals that, even on a small knowledge database with 6 million entries only, regular retrieval average incurs 0.62 seconds latency\footnote{Tested with \texttt{Faiss}-\texttt{IndexFlat} on our workstation.}. While it is far from the reality scale, the observed latency exceeds the TTFT of bare LLMs by a large margin\footnote{Data source: https://github.com/tenstorrent/tt-metal}. The rise of agentic RAG pipelines further intensifies this challenge, as iterative retrieval over decomposed sub-queries amplifies the overhead \cite{yu2024b, liu2024b, yang2024a, gao2024e}.

With respect to retrieval acceleration, existing approaches generally fall into two paradigms. Approximate nearest neighborhood search (ANNS), such as IVF and ScaNN \cite{scann}, are well-established in industry, as shown in Fig.\ref{fig:anns}. They partition the database into buckets and retrieve only from the closest ones. The reduced retrieval scope accelerates retrieval, but inevitably compromises accuracy \cite{quinn2025}. In contrast, reuse-based methods \cite{frieder2024, bergman2025} aim to reuse retrieval results of recent queries with strictly identical semantics, as illustrated in Fig.\ref{fig:reuse}. Due to the diversity of the real world, such a query infrequently exists, leading to marginal efficiency gains.

Inspired by the speculative decoding in LLM, we propose \textbf{HaS}, a \textbf{H}omology-\textbf{a}ware \textbf{S}peculative Retrieval method. HaS is designed as a plug-and-play service that intercepts queries prior to full-database retrieval. Conceptually, the framework first executes rapid retrieval over a restricted subset to generate a "draft" of candidate documents. It then leverages the homology relation between queries to validate whether this draft contains the necessary knowledge to be accepted. High-latency full-database retrieval is thus bypassed whenever the draft is accepted.

Specifically, as depicted in Fig.\ref{fig:speculative}, HaS first executes \textbf{two-channel fast retrieval} to speculatively obtain a draft. Both channels maintain a small retrieval scope to ensure low latency. The cache channel comprises documents retrieved from recent queries, providing initial knowledge coverage while serving as probes for the subsequent validation stage. The fuzzy channel maintains an extremely narrow subset of the database, refining draft quality and enhancing validation reliability by capturing knowledge that may be absent from the cache.

The core of HaS is the \textbf{homology validation} mechanism, which determines whether a draft can be accepted. Our acceptance principle is that the draft contains the right document(s) to support a factual response. Direct verification via query-document semantic evaluation is computationally prohibitive. Instead, we introduce a symmetric query–query relation termed \textit{homology}, which captures entity alignment between queries. By leveraging this relation, the validation is reduced to determining whether a previously cached query constitutes a homologous re-encounter of the incoming query. A failure to re-identify provides strong evidence for rejecting the draft, whereas successful re-identification justifies its acceptance. Given the high frequency of homologous queries in real-world popularity patterns \cite{agarwal2025}, HaS has huge potential for accelerating RAG.

\begin{figure}[htbp]
  \centering
  \subfloat[Approximate Nearest Neighbor Search \label{fig:anns}]{
    \includegraphics[width=0.8\linewidth]{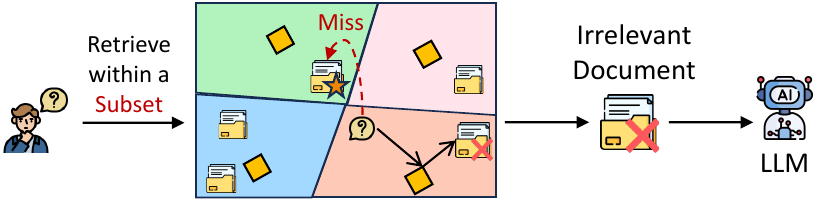}
  }
  \hfill
  \subfloat[Reuse-based Retrieval\label{fig:reuse}]{
    \includegraphics[width=0.8\linewidth]{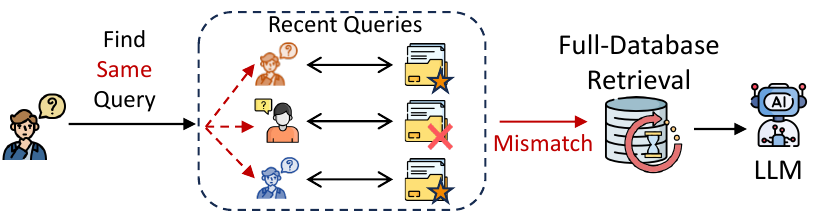}
  }
  \hfill
  \subfloat[Homology-aware Speculative Retrieval (HaS) \label{fig:speculative}]{
    \includegraphics[width=0.8\linewidth]{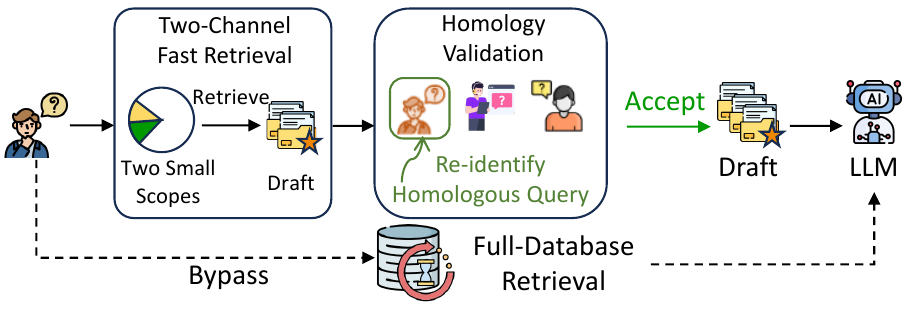}
  }
  \caption{Illustration of different approaches for accelerating retrieval in RAG.}
  \label{fig:main}
\end{figure}

Extensive experiments demonstrate that, when integrated into a standard RAG pipeline, HaS reduces retrieval latency by 23.74\% and 36.99\% across datasets, while incurring only a marginal 1–2\% degradation in response accuracy. It delivers superior performance compared to competitive methods. HaS can also be seamlessly integrated into modern agentic RAG pipelines featuring query decomposition and iterative retrieval, yielding more substantial end-to-end latency reductions.

The contributions of this work are summarized as follows:
\begin{enumerate}
    \item We design HaS to accelerate retrieval in RAG. To conditionally bypass slow full-database retrieval, HaS speculatively retrieves from two channels to quickly obtain a draft, followed by a validation procedure.
    \item The charm of HaS lies in its homology validation. By re-identifying homologous queries from the cache as evidence for draft acceptance, HaS achieves both high efficiency and accuracy.
    \item Experiments show that HaS cuts retrieval latency by 23.74\% and 36.99\% across datasets in the standard RAG pipeline, outperforming SOTA methods. Being pluggable to modern RAG pipelines, this benefit becomes even more significant.
\end{enumerate}

\begin{figure*}[htbp]
  \centering
  \includegraphics[width=\linewidth]{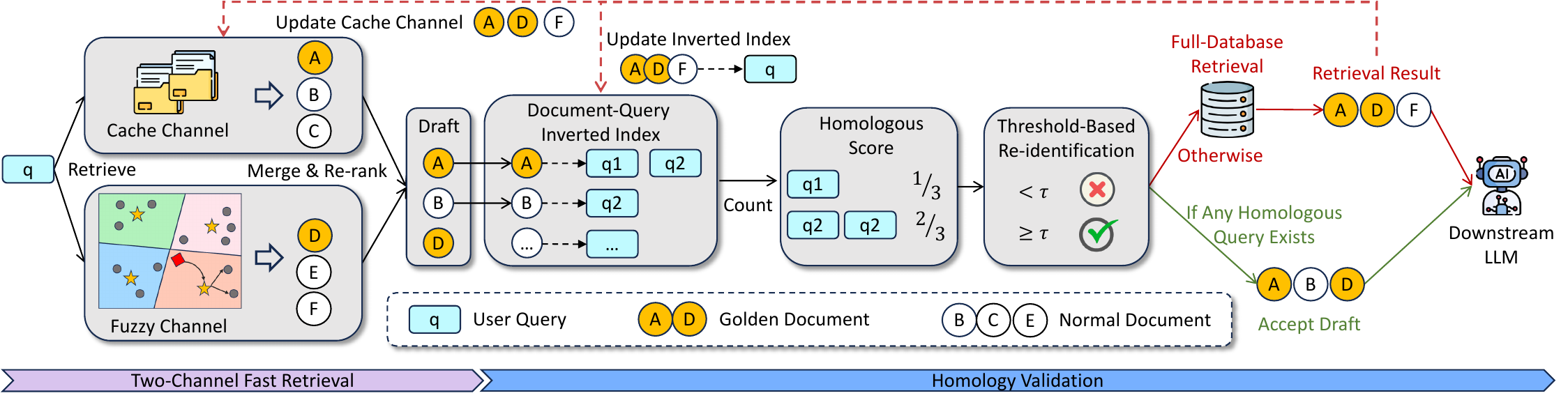}
  \caption{Framework of HaS. Given a query $q$, the two-channel fast retrieval is first performed. Documents A–F are retrieved, and the Top-3 (A, B, and D) form the draft. For validation, documents in the draft are indexed to cached queries by the document-query inverted index. Frequencies of queries hit in the cache are used to compute their homology scores for threshold-based re-identification. If any quasi-homologous query is re-identified, the draft is accepted; otherwise, a full-database retrieval is performed, and results are used to update the cache channel and the inverted index.}
  \label{fig:framework}
\end{figure*}

\section{The Overall Pipeline of HaS}
In this section, we present the problem formulation and provide an overview of HaS. The design rationale is presented in the next section. Key notations are summarized in Table \ref{tab:notation}

\begin{table}[t]
\centering
\caption{Key Notations}
\begin{tabular}{ll}
\hline
\textbf{Symbol} & \textbf{Description} \\
\hline
$q$ & Current user query \\
$d$ & Document \\
$E(\cdot)$ & Target entity \\
$A(\cdot),\,\mathcal{A}(\cdot)$ & Target attribute(s) \\
$\mathcal{D}$ & A draft as set of documents retrieved \\
$\mathcal{C}_c,\;\mathcal{C}_f$ & The cache channel and fuzzy channel \\
$\mathcal{P}=\{(q_h,\mathcal{D}_h)\}$ & Cached queries and retrieved documents \\
$H_{\max}$ & Cache capacity \\
$s(\cdot)$ & Homology score \\
$\mathcal{J}$ & Inverted index mapping documents to cached queries \\
$k$ & Number of documents to retrieve \\
$\tau$ & Threshold for homology validation \\
$G(d,q)$ & Indicator that $d$ is a golden document for $q$ \\
$H(\cdot,\cdot), H^Q(\cdot,\cdot)$ & Indicators of homology and quasi-homology\\
$\mathcal{V}(\mathcal{D},q), \hat{\mathcal{V}}(\mathcal{D},q)$ & Draft acceptance hypothesis and its surrogate \\
\hline
\end{tabular}
\label{tab:notation}
\end{table}

\subsection{Problem Formulation}
In a bare RAG pipeline, the system maintains a large-scale corpus $\mathcal{C}$. Each document $d \in \mathcal{C}$ contains information regarding several attributes $\mathcal{A}(d)$ of an entity $E(d)$. Given a query $q$, the retrieval stage recalls a set of relevant documents $\hat{\mathcal{D}}$, which are prepended to the LLM's prompt to generate a response $L(q,\hat{\mathcal{D}})$.

We focus on accelerating the retrieval stage. We assume that queries are well-organized, targeting a specific attribute $A(q)$ of a target entity $E(q)$. For complex, multi-hop queries, we can follow the standard practice of performing query decomposition prior to retrieval \cite{liu2024b,yu2024b} (an use case is provided in Section \ref{use case}).

For each query $q$, HaS first performs retrieval over a restricted scope $\mathcal{C}_s \subset \mathcal{C}$ to obtain a $k$-document draft $\mathcal{D}$ with an average latency of $\bar{\ell}_s$. Since retrieval latency scales with the database size, we have $\bar{\ell}_s \ll \bar{\ell}$, where $\bar{\ell}$ is the average retrieval latency on $\mathcal{C}$. We define a validation function $R(q,\mathcal{D})\in\{0,1\}$, where $R(q,\mathcal{D})=1$ indicates that the draft is acceptable and vice versa (the validation mechanism is detailed in Section \ref{sec3}). If acceptable, $\mathcal{D}$ is returned; otherwise, full-database retrieval is invoked to obtain $\mathcal{D}_{\text{full}}$. The final result set is therefore:
\begin{equation}
    \hat{\mathcal{D}} =
        \begin{cases}
        \mathcal{D}, & R(q,\mathcal{D})=1,\\
        \mathcal{D}_{\text{full}}, & R(q,\mathcal{D})=0.
        \end{cases}
\end{equation}

By optimizing $\mathcal{C}_s$ and $R$, HaS aims to reduce retrieval latency while maximizing response accuracy. For a query set $\mathcal{Q}$, the multi-objective optimization problem is formulated as:

\begin{equation}
    \min_{\mathcal{C}_s, R} \frac{1}{|\mathcal{Q}|}
    \sum_{q\in\mathcal{Q}}
    \Big(
        R(q,\mathcal{D})\,\bar{\ell}_s +
        (1-R(q,\mathcal{D}))(\bar{\ell}_s+\bar{\ell})
    \Big),
\nonumber
\end{equation}
\begin{equation}
    \max_{\mathcal{C}_s, R} \frac{1}{|\mathcal{Q}|} \sum_{q\in\mathcal{Q}}
    Acc(L(q,\hat{\mathcal{D}})),
    \nonumber
\end{equation}
where $Acc(\cdot) \in \{0, 1\}$ measures whether the response matches the reference answer.

\subsection{Overview of HaS}
HaS is designed as a pluggable, lightweight component before the full-database retrieval, as illustrated in Fig.\ref{fig:framework}. Following modern service practice, such as Content Delivery Networks (CDNs), it can be deployed proximate to the LLM to enable low-latency access. Incoming queries can be first processed by HaS in a local network, while those with draft rejection are redirected to the full database hosted remotely. The pseudo-code of HaS is provided in Algorithm \ref{alg:has}.

HaS maintains two retrieval channels whose union constitutes $\mathcal{C}_s$, namely the cache channel and the fuzzy channel. The combined retrieval scope of these channels is restricted for fast retrieval. The cache channel is populated by documents previously retrieved for historical queries. Formally, let $\mathcal{P} = \{(q_h, \mathcal{D}_h) \mid 0 \leq h \leq H_{\text{max}}\}$ represent the system cache, where each entry consists of a historical query $q_h$ and its corresponding full-database retrieval result set $\mathcal{D}_h$, up to a maximum capacity $H_{\text{max}}$. The cache channel $\mathcal{C}_c = \bigcup_{h=0}^{H_{\text{max}}} \mathcal{D}_h$ is defined as the union of all documents within these cached results\footnote{Without ambiguity, the cache $\mathcal{P}$ denotes the query–documents pairs, while the cache channel $\mathcal{C}_c$ refers to the collection of all documents contained in the cache.}. The fuzzy channel is implemented by ANNS, but configured aggressively to retrieve only a very narrow subset. This allows it to return results rapidly, albeit with reduced accuracy. The benefits of introducing these two channels are detailed in Section \ref{why fuzzy source}.

For each query $q$, HaS first executes two-channel fast retrieval. Top-$k$ documents from both channels are recalled, denoted as $\mathcal{D}_c$ and $\mathcal{D}_f$, where $\mathcal{D}_i = \arg\operatorname{top}^k_{d \in \mathcal{C}_i} \text{sim}(g(q), g(d)), i \in \{c, f\}$, and $g(\cdot)$ is an semantic encoder. These documents are re-ranked, and the top-$k$ documents are selected as the draft $\mathcal{D}=\arg\operatorname{top}^k_{d\in\mathcal{D}_c\cup\mathcal{D}_f} \text{sim}(g(q), g(d))$.

After the draft is obtained, HaS initiates homology validation by seeking to re-identify a homologous counterpart in the cache, who share the same target entity of the current query. This relationship is quantified by the homology score, defined as the overlap ratio between the retrieval result sets of the two queries. Queries whose homology score is higher than a preset threshold are considered homologous. The formal definition of homology is given in Section \ref{What Homology is}, while the mechanism of draft validation will be elaborated in Sections \ref{From Validation to Re-identification} and \ref{Re-identification via Homology Score}.

To efficiently compute homology scores against all cached queries, HaS maintains a document-query inverted index $\mathcal{J}: \mathcal{D} \rightarrow \mathcal{Q}$, which maps each document $d$ to a set of cached queries whose retrieval results include $d$, i.e. $\mathcal{J}(d) = \{ q_h \mid (q_h, \mathcal{D}_h) \in \mathcal{P}, d \in \mathcal{D}_h \}$. During validation, $\mathcal{J}$ indexes each $d \in \mathcal{D}$ to its associated cached queries, forming a multiset $\mathcal{M} = \bigcup_{d \in \mathcal{D}} \mathcal{J}(d)$. For each cached query $q_h$, its homology score is thus computed as $s(q_h) = f(q_h) / k$, where $f(q_h)$ denotes the frequency of $q_h$ in $\mathcal{M}$. If there exists a $q_h$ such that $s(q_h) > \tau$, the query is considered homologous and the draft is accepted. Otherwise, HaS falls back to full-database retrieval, and correspondingly updates $\mathcal{P}$, $\mathcal{C}_c$, and $\mathcal{J}$.

\begin{algorithm}
\caption{HaS}
\label{alg:has}
\begin{algorithmic}[1]

\Statex \textbf{Input:} A user query $q$.
\Statex \textbf{Output:} A set of documents.

\State Perform two-channel fast retrieval from both $\mathcal{C}_f$ and $\mathcal{C}_c$ to obtain candidate sets $\mathcal{D}_f$ and $\mathcal{D}_c$.
\State Re-rank and merge candidates to form the draft $\mathcal{D}$.

\State Initialize a multiset $\mathcal{M} = \emptyset$.
\For {each $d \in \mathcal{D}$}
    \State Find queries through the inverted index $\mathcal{J}(d)$.
    \State $\mathcal{M} = \mathcal{M} \cup \mathcal{J}(d)$
\EndFor

\For {each unique query $q_h \in \mathcal{M}$}
    \State Count its occurrence frequency $f(q_h)$.
    \State Compute the homology score: $s(q_h)=f(q_h) / k$.
    \If {$f(q_h) / k > \tau$}
        \State \Return Accept draft $\mathcal{D}$.
    \EndIf
\EndFor

\State Perform full-database retrieval to obtain result $\mathcal{D}_{\text{full}}$.
\State Update $\mathcal{P}$, $\mathcal{C}_c$ and $\mathcal{J}$.
\State \Return $\mathcal{D}_{\text{full}}$

\end{algorithmic}
\end{algorithm}

\section{Key Mechanisms in HaS}
\label{sec3}
In this section, we introduce the key mechanisms of HaS by explaining: (1) what homology is, (2) how validation is formulated as a homologous query re-identification problem, (3) how homology is quantified for re-identification, and (4) why two channels are incorporated into the framework.

\subsection{What Homology is}
\label{What Homology is}
To guarantee factuality, the recalled set of documents in the retrieval stage must include at least one document to provide the necessary supporting evidence. In common practice, this is judged by the alignment of entities and attributes between the query and the document, which forms the basis for our definition of \textit{golden documents}:
\vspace{0.5\baselineskip}
\begin{definition}[Golden document]
    A document $d$ is a \textbf{golden document} for a query $q$ if it contains the evidence to support a factual response. Concretely, $d$ should correspond to the target entity and cover the specific attribute of interest. We define the indicator function:
    \[
    G(d, q) \triangleq \mathbb{I}\left[ E(q) = E(d) \right] \land \mathbb{I}\left[ A(q) \in \mathcal{A}(d) \right],
    \]
    where $\mathbb{I}(\cdot)$ equals $\text{true}$ if the condition holds and false otherwise. $d$ is a golden document to $q$ if $G(d, q)=\text{true}$.
\end{definition}
\vspace{0.5\baselineskip}

This definition implies a fundamental relationship between queries: when two queries align in both target entity and attributes, they necessarily share an overlapping set of golden documents. We characterize this relationship as \textit{full homology}, as illustrated in Fig.\ref{fig:definition_fully_homo}:

\vspace{0.5\baselineskip}
\begin{definition}[Full Homology]
\label{fully homologous query}
    \textbf{Full Homology} exists between two queries $q_1$ and $q_2$, or the two queries are considered \textbf{fully homologous} if they point to the same entity and cover the same attributes of interest:
    \[
        H_f(q_1, q_2) \triangleq \mathbb{I}\big[E(q_1) = E(q_2)\big] \land \mathbb{I}\big[A(q_1) = A(q_2)\big].
    \]
\end{definition}
\vspace{0.5\baselineskip}

Existing reuse-based methods leverage this golden-document sharing property by matching fully homologous cached queries, which typically require identical semantics, to facilitate result reuse. However, their potential for acceleration is severely constrained in practice, as the inherent diversity of real-world user queries makes such strict semantic equivalence exceptionally rare.

\begin{figure}[htbp!]
  \centering
  \includegraphics[width=0.9\linewidth]{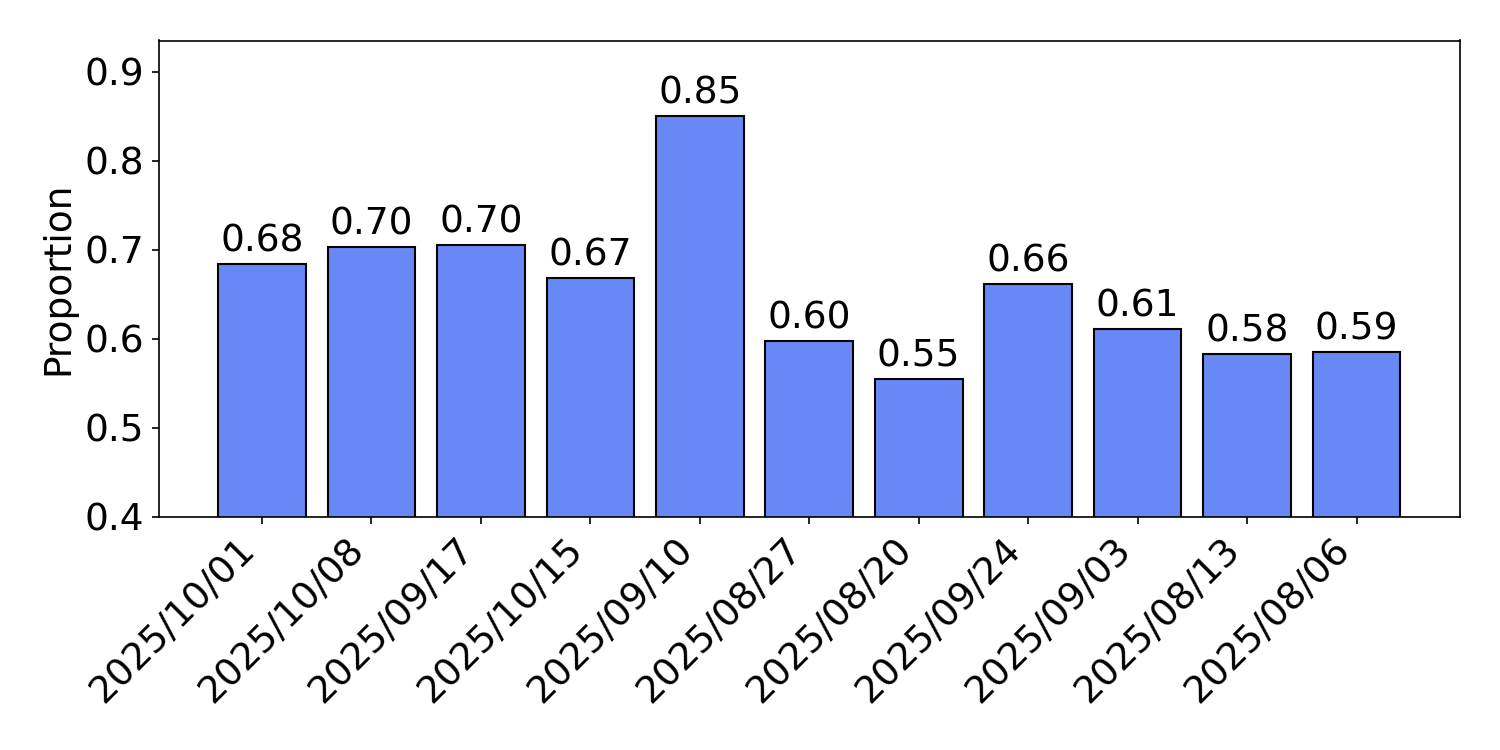}
  \caption{Estimated proportion of queries that have homologous counterparts in the real world.}
  \label{fig:proportion_of_homo_queries}
\end{figure}

This diversity is primarily driven by varying attribute targets, while real-world queries nonetheless follow the popularity pattern to cluster around popular entities. Pioneering analysis of search engine logs reveals that entities are queried 3.97 times on average, with over 83.9\% of queries sharing a target entity with at least one other request. 
We conduct a similar analysis using Wikipedia’s TopViews statistics. Specifically, we curated a dataset of the Top-1000 most-visited Wikipedia entities, sampled weekly every Wednesday over a three-month duration\footnote{Data Source: \url{https://pageviews.wmcloud.org/topviews}. View counts are normalized relative to the least-visited page. Entities with a normalized count exceeding 2 are identified as having at least one homologous counterpart.}. As illustrated in Fig.\ref{fig:proportion_of_homo_queries}, over 60\% of queries have such counterparts, and the proportion rises remarkably during major or trending events (e.g., “Charlie Kirk” on September 10, 2025). We refer to this special relationship between queries as \textit{homology}, as shown in Fig.\ref{fig:definition_partial_homo}:
\vspace{0.5\baselineskip}
\begin{definition}[Homology]
\label{homologous query}
    \textbf{Homology} exists between two queries $q_1$ and $q_2$, or the two queries are considered \textbf{homologous} if they point to the same entity:
    \[
        H(q_1, q_2) \triangleq \mathbb{I}\big[E(q_1) = E(q_2)\big].
    \]
\end{definition}
\vspace{0.5\baselineskip}

In line with full homology, our empirical analysis shows that homologous queries tend to share a subset of golden documents. Based on this observation, we first formalize the weaker form of golden-document sharing as \textit{quasi-homology}:
\vspace{0.5\baselineskip}
\begin{definition}[Quasi-homology]
\label{quasi-homologous query}
    We define \textbf{quasi-homology} between two queries $q_1$ and $q_2$ if they share at least one golden document:
     \[
        H^Q(q_1, q_2) \triangleq \mathbb{I} \big[ \exists d, G(d, q_1) = G(d, q_2) = \text{true} \big].
     \]
\end{definition}
\vspace{0.5\baselineskip}

We then bridge these two concepts through the following insight:
\vspace{0.5\baselineskip}
\begin{insight}
    Given two homologous queries $q_1, q_2$, they are empirically quasi-homologous.
\end{insight}
\vspace{0.5\baselineskip}

This empirical insight is substantiated by the following two observations:

\textbf{1) Retrieval Exhibits Strong Entity Alignment.} Analysis of retrieval results reveals that an average of 2.35 out of the top-5 documents are entity-aligned with the query, with 64.3\% of queries having an entity-aligned document in the top-1 position. This alignment is primarily attributed to an entity-centric bias within semantic encoders, as positive training samples often contain the same entity. Consequently, a golden document $d$ for query $q_1$ inherently satisfies the condition $E(d) = E(q_1) = E(q_2)$.

\textbf{2) Documents Provide Multi-Attribute Coverage.} Comprehensive knowledge sources, such as Wikipedia, typically describe multiple attributes of an entity within a single document. As reported in \cite{agarwal2025}, 5\% of documents fulfill 60\% of queries, demonstrating the rich knowledge coverage of documents to support distinct queries. This implies a high probability that a document $d$ supporting $q_1$ also encompasses the specific attribute $A(q_2)$ required by query $q_2$, i.e., $A(q_2) \in \mathcal{A}(d)$.

Collectively, these observations imply that two homologous queries, $q_1$ and $q_2$, are highly likely to share at least one golden document, thereby exhibiting quasi-homology.

In the subsequent sections, we leverage this prevalent homologous relationship, alongside this insight, to develop a robust and efficient mechanism for draft validation.

\begin{figure}[htbp]
  \centering
  \subfloat[Full Homology\label{fig:definition_fully_homo}]{
    \includegraphics[width=0.9\linewidth]{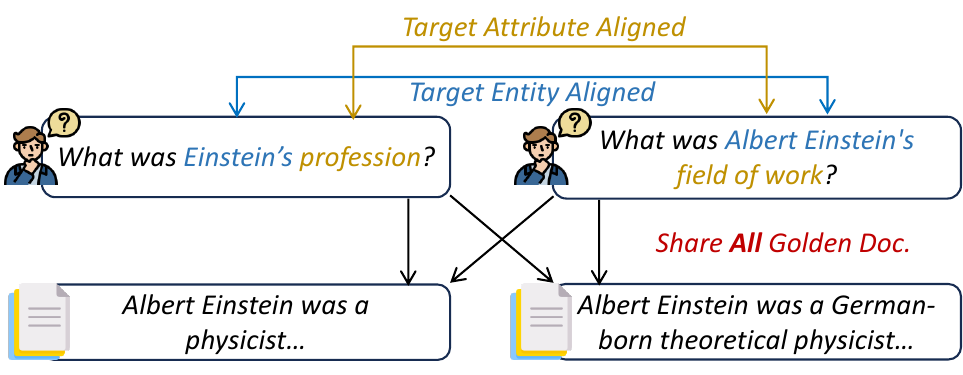}
  }
  \hfill
  \subfloat[Homology\label{fig:definition_partial_homo}]{
    \includegraphics[width=0.9\linewidth]{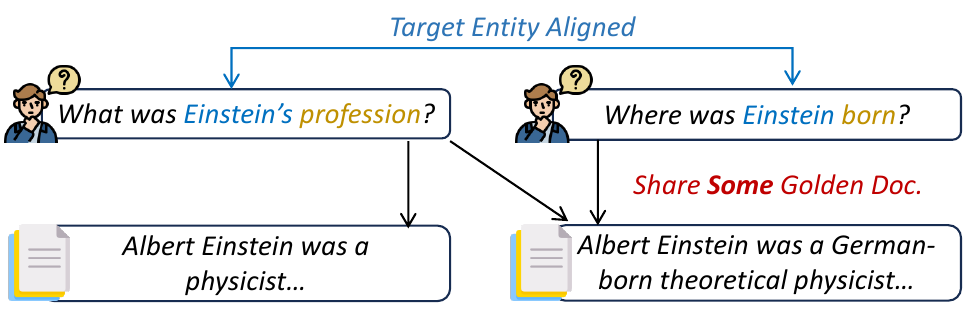}
  }
  \caption{An illustration of (fully) homologous queries.}
  \label{fig:definition of homo}
\end{figure}

\subsection{From Validation to Re-identification}
\label{From Validation to Re-identification}
For candidate documents (i.e., a draft) speculatively retrieved, we need to validate whether it is acceptable following the rule: \textbf{a draft $\mathcal{D}$ for query $q$ is considered acceptable if it contains at least one golden document}. We express it as the following hypothesis:
\[
\mathcal{V}(\mathcal{D},q): \exists d^*\in \mathcal{D} \; \text{s.t.} \; G(d^*,q)=\text{true}.
\]
Thus, \textbf{the draft validation process is to verify whether $\mathcal{V}(\mathcal{D}, q)$ holds.} If it is true, the draft can be accepted.

Since both $q$ and $d$ are in natural language, directly verifying $\mathcal{V}(\mathcal{D}, q)$ requires a reliable evaluator (e.g., an LLM) for $G(d, q)$, which is very costly. Hence, we instead introduce a surrogate hypothesis $\hat{\mathcal{V}}(\mathcal{D},q)$. It brings in a reference query $q_h$ in the cache $\mathcal{P}$, requiring that $d^*$ is a golden document for both $q$ and $q_h$ and appears in the cached retrieval result $\mathcal{D}_h$:
\[
\begin{aligned}
    \hat{\mathcal{V}}(\mathcal{D},q): \; & \exists (q_h, \mathcal{D}_h) \in \mathcal{P}, \exists d^* \in \mathcal{D} \cap \mathcal{D}_h, \\
         \text{s.t.} \; & G(d^*, q) = G(d^*, q_h) = \text{true}. \\
\end{aligned}
\]

This surrogate hypothesis is not overly strict. Under our two-channel design, the recalled document $d^*$ is likely drawn from the cache channel, meaning it belongs to some $\mathcal{D}_h$ of a cached query $q_h$. Meanwhile, since $\mathcal{D}_h$ is obtained through full-database retrieval, $d^*$ is likely golden to $q_h$. In this way, \textbf{validating $\hat{\mathcal{V}}(\mathcal{D},q)$ equivalently verifies $\mathcal{V}(\mathcal{D},q)$}.

The existence of such a $q_h$ in the cache provides key evidence for validating the surrogate hypothesis $\hat{\mathcal{V}}(\mathcal{D}, q)$. Recall that quasi-homologous queries share at least one golden document, and that homologous queries are likely to exhibit quasi-homology. Therefore, if the surrogate hypothesis $\hat{\mathcal{V}}(\mathcal{D}, q)$ holds, the presence of the shared golden document $d^*$ implies that $q_h$ and $q$ must be quasi-homologous, and consequently homologous. Formally, we have:
\begin{equation}
    \hat{\mathcal{V}}(\mathcal{D},q) \implies \exists q_h\; \text{s.t.}\;H^Q(q, q_h) \implies \exists q_h\; \text{s.t.}\;H(q, q_h).
    \nonumber
\end{equation}

\begin{figure*}[htb!]
  \centering
  \subfloat[\label{fig:distributions of scores, easy positive}]{
    \includegraphics[width=0.31\linewidth]{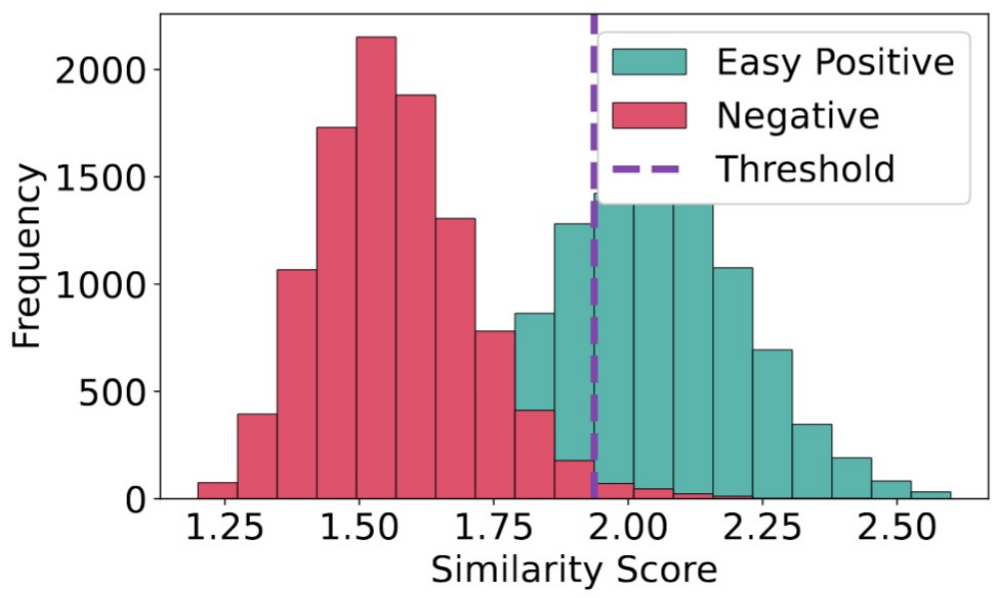}
  }
  \hfill
  \subfloat[\label{fig:distributions of scores, hard positive, semantic score}]{
    \includegraphics[width=0.31\linewidth]{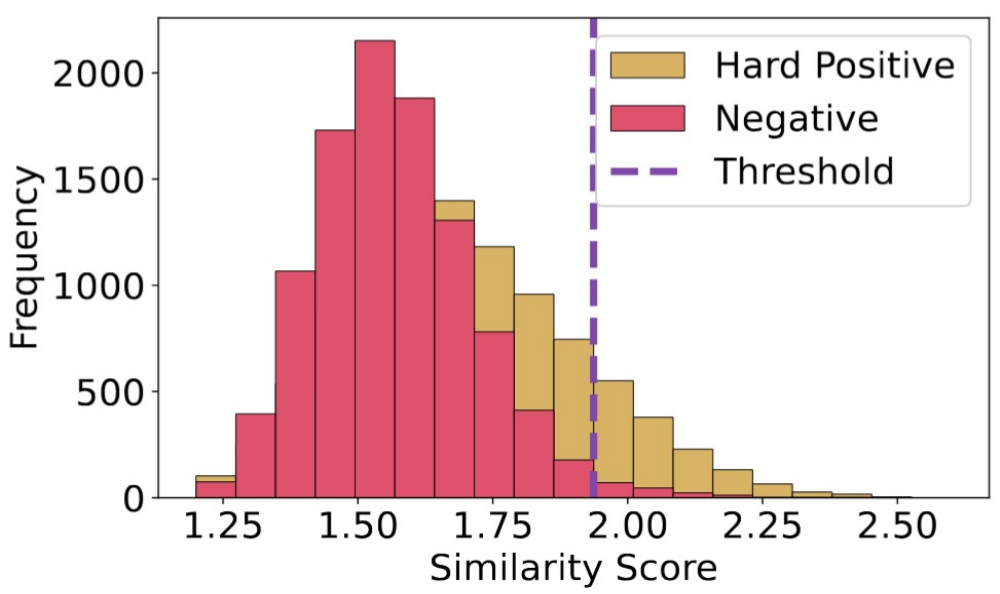}
  }
  \hfill
  \subfloat[\label{fig:distributions of scores, hard positive, homology score}]{
    \includegraphics[width=0.31\linewidth]{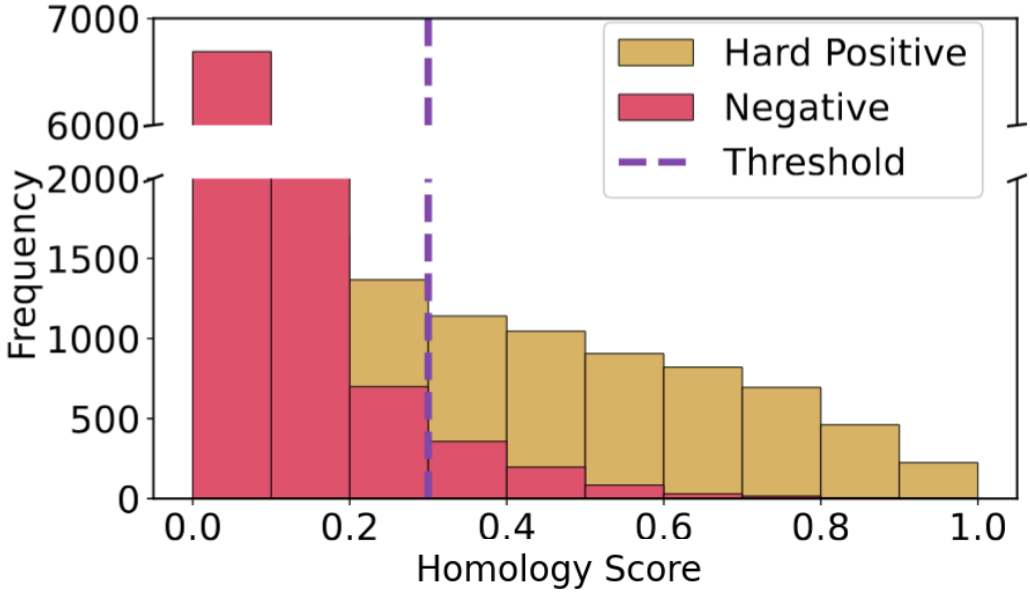}
  }
  \caption{Distributions of semantic similarity scores and homology scores for Easy Positives (Fully homologous), Hard Positives (Homologous), and Negatives (Other) queries. A larger positive area beyond the threshold indicates that more required queries can be distinguished from others to support draft validation.}
  \label{fig:distributions of scores}
\end{figure*}

More importantly, the contrapositive shows that, if none of such a $q_h$ exists, the hypothesis is directly proved false:
\begin{equation}
\label{contrapositive}
    \forall q_h\;\text{s.t.}\;\neg H(q, q_h) \implies \neg\hat{\mathcal{V}}(\mathcal{D},q),
    \nonumber
\end{equation}
which helps the system reject the draft with confidence.

Drawing an analogy to person re-identification, we formulate draft validation as a homologous query re-identification problem. Specifically, the objective is to determine whether an incoming query constitutes a homologous re-encounter of a previously observed query, or equivalently, whether any cached query is homologous to the current one. If no such homologous counterpart can be identified within the cache, the draft is rejected with high confidence; otherwise, it is deemed acceptable.

\subsection{Re-identification via Homology Score}
\label{Re-identification via Homology Score}
Efficient and accurate re-identification of homologous queries presents a challenge. Conventional lexical matching struggles with entity alignment due to the inherent issues of linguistic ambiguity and polysemy. While LLMs can enhance robustness, their inference latency is prohibitive. Furthermore, the semantic similarity score, which is frequently employed to identify fully homologous queries, is insufficient for this task. As demonstrated in Fig.\ref{fig:distributions of scores, easy positive} and Fig.\ref{fig:distributions of scores, hard positive, semantic score}, homologous queries are nearly indistinguishable from non-homologous ones, since semantics are meanwhile influenced by attribute variations.

Recalling the entity-alignment property inherent in retrieval, entity-aligned homologous queries are predisposed to recall overlapping sets of entity-related documents. Building upon this observation, we exploit the overlap ratio between retrieval results as a metric for the homology relation between queries. We formally define the \textit{homology score} to quantify the likelihood that a reference query is homologous to a target query:
\vspace{0.5\baselineskip}
\begin{definition}[Homology score]
\label{h_score}
    Given a query $q_1$ and a reference query $q_2$, let $\mathcal{D}_1$ and $\mathcal{D}_2$ denote their respective retrieval results containing $k$ documents each. The \textbf{homology score} is defined as the proportion of overlapping documents between the two sets: $s(q_1, q_2) = |\mathcal{D}_1 \cap \mathcal{D}_2| / k$.
\end{definition}
\vspace{0.5\baselineskip}

Fig.\ref{fig:distributions of scores, hard positive, homology score} visualizes the efficacy of the homology score. With the score, homologous queries can be easily separated from the others with a noticeably distinct distributional gap. Quantitatively, the recall reaches 53.28\%, significantly outperforming the semantic similarity score.

With this metric, a threshold-based re-identification process can be performed, where we try to find a cached query that happens to have a high homology score with the given query. If no cached query satisfies this criterion, the draft is rejected with confidence. Otherwise, the existence of a homologous query $q_h$ to $q$ in the cache is established. A high overlap between $\mathcal{D}$ and $\mathcal{D}_h$ indicates a greater likelihood that their shared golden document $d^*$ lies in $\mathcal{D} \cap \mathcal{D}_h$, thereby strengthening the surrogate hypothesis $\hat{\mathcal{V}}(\mathcal{D},q)$ and supporting draft acceptance.

Since we have past queries and their documents cached, instead of computing the homology score for each query, we leverage a document-query inverted index $\mathcal{J}(d):\mathcal{D}\rightarrow \mathcal{Q}$ for acceleration. For any draft $\mathcal{D}$ to be validated, each retrieved document acts as a probe that maps back to a set of cached queries, yielding a multiset $\mathcal{M_D} = \bigcup_{d \in \mathcal{D}} \mathcal{J}(d)$. In this way, the frequency $f(q_h)$ of query $q_h$ in $\mathcal{M}$ reflects the number of overlapping documents between its retrieval result and the draft. The homology score $s(q, q_h)$ can thus be computed by $f(q_h) / k$. If $\max_{q_h} s(q, q_h) > \tau$, the draft can be accepted.

\subsection{Why two channels are incorporated}
\label{why fuzzy source}

The caching mechanism in HaS is designed to sustain robust homology validation and preserve adequate knowledge coverage, while still enabling fast retrieval. To achieve this, HaS integrates two distinct channels: the cache channel and the fuzzy channel.

\textbf{The cache channel, which consists of the retrieved documents of cached queries, is an indispensable component.} It provides the most important knowledge coverage. Once a homologous query is re-identified, its associated documents in the cache channel are likely golden to the current query, supporting a correct response. These documents also serve as probes to support homology validation. When some of them appear in the draft, they are linked back to the corresponding cached queries for re-identification.

However, using the cache channel alone faces two critical issues. First, ranking-based retrieval introduces noisy documents. When only $<k$ relevant documents exist, the Top-$k$ retrieval result is filled with noisy ones. These noisy documents inflate the homology scores of non-homologous queries and potentially lead to incorrect re-identification. Second, the cache spans only a small portion of the knowledge space and often lacks fine-grained information specific to the target attribute. To address these limitations, we introduce the fuzzy channel as a complement.

\textbf{The fuzzy channel restricts retrieval to a very small subset, returning fuzzy results with extremely low latency.} We implement it using ANNS with an aggressive configuration. We partition the database into thousands of buckets, and retrieve only a few dozen of the closest ones. This yields fuzzy yet sufficiently informative documents. We deliberately accept the associated moderate accuracy loss in exchange for extremely fast retrieval. Retrieved documents from both channels are merged and re-ranked to form the draft.

\textbf{Incorporating the fuzzy channel enhances both validation reliability and draft quality.} First, although the fuzzy channel produces fuzzy results, after re-ranking, these weakly-relevant documents can edge out noisy documents retrieved from the cache channel in the draft. This mitigates noise-induced homology score inflation, thereby avoiding incorrect re-identification and improving validation reliability. Second, retrieval from the closest buckets improves alignment with the attribute-specific intent of queries, enabling the fuzzy channel to capture fine-grained knowledge that may not exist in the cache channel. Meanwhile, documents from both channels compete for inclusion in the draft. Only highly relevant ones are retained, while noisy candidates are discarded. Thus, the two-channel design yields a high-quality and factual draft. An illustrative example is provided in Section \ref{use case} to clarify the above discussion.

\section{Experiments}
\subsection{Experimental Settings}

\textbf{Implementation Details.} We implement HaS on a workstation equipped with an I9-13900KF CPU and an RTX-3090 GPU. We primarily use $\texttt{Qwen3}_{8B}$ \cite{qwen3technicalreport} as the downstream LLM, and additionally evaluate $\texttt{Llama3}_{8B}$ \cite{llama3modelcard} and $\texttt{Mixtral}_{7B}$ \cite{jiang2023mistral7b}, all in their AWQ-quantized versions. We simulate a cloud–edge system for deployment. The full-dataset retrieval service is deployed on the cloud, with a simulated network latency of 0.1–0.2 seconds injected. It is implemented with \texttt{Faiss-IndexPQ} \cite{douze2024faiss}. HaS are simulated on the edge with a network latency of 0.01-0.05 seconds. The fuzzy channel is implemented by \texttt{Faiss-IndexIVFPQ} to retrieve 64 out of 8192 buckets. The cache channel uses \texttt{HNSWLib}. A maximum of 5000 queries with their retrieved documents can be cached. A first-in-first-out (FIFO) policy is applied for cache replacement. Cold start is always included. In default, $k$ is set to 10 and $\tau$ is set to 0.2.

\textbf{Dataset Selection and Augmentation.}
We collect the \texttt{Wikipedia documents dump} on 2023-11-01\footnote{\url{https://huggingface.co/datasets/wikimedia/wikipedia}}, and segment it into 49.2 million passages of less than 100 words \cite{karpukhin-etal-2020-dense}. 

To simulate realistic query patterns, we first note that most existing QA datasets are filtered and de-duplicated. For example, in the Granola-EQ dataset, most entities appear in only a single query, as shown in Fig.\ref{fig:frequency comparison, granola, origin}. These datasets overlook the tendency of real-world queries to concentrate on popular entities, leading to a deviation from realistic usage patterns.

\begin{figure}[htbp]
  \centering
  \subfloat[Granola-EQ (Original)\label{fig:frequency comparison, granola, origin}]{
    \includegraphics[width=0.47\linewidth]{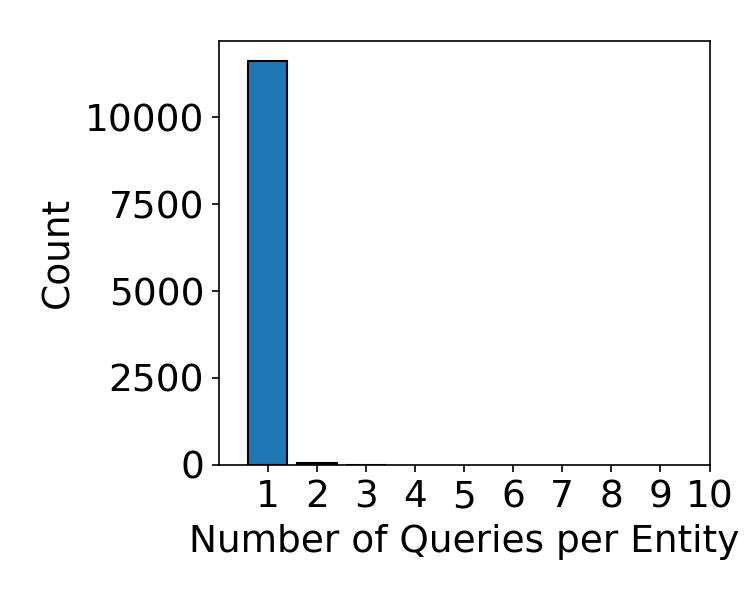}
  }
  \hfill
  \subfloat[Granola-EQ (Augmented)\label{fig:frequency comparison, granola, sampled}]{
    \includegraphics[width=0.47\linewidth]{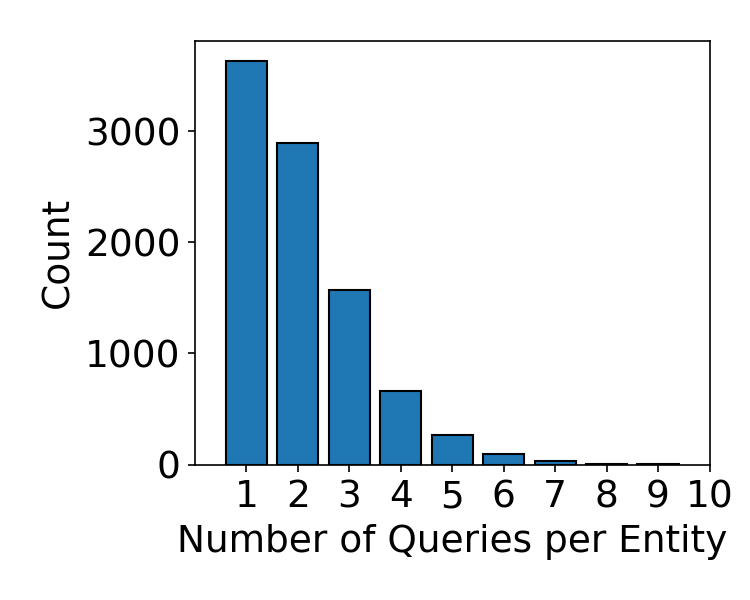}
  }
  \caption{Distribution of the number of attributes queried per entity. After sampling queries from the augmented dataset, the resulting distribution better reflects real-world popularity patterns.}
  \label{fig:query distribution}
\end{figure}

To address this limitation, we augment existing datasets by leveraging Wikidata, a knowledge graph that connects entities through diverse relations. We primarily use the \texttt{Granola-EQ} dataset \cite{yona2024narrowing}, which augments the classical EQ dataset \cite{sciavolino2021simple} with entity annotations, and also include \texttt{PopQA} \cite{popqa}. The workflow is illustrated in Fig.\ref{fig:dataset_augment}. We leverage their annotated entities to find relations and linked entities. A subset of representative relations is selected. For each relation, we create 5 templates and fill them to obtain the question–answer pairs. To better reflect real-world patterns, we sample queries from the augmented dataset while aligning with the daily TopView Statistics of Wikipedia, as illustrated in Fig.\ref{fig:frequency comparison, granola, sampled}.

\begin{figure}[htbp]
  \centering
  \includegraphics[width=0.9\linewidth]{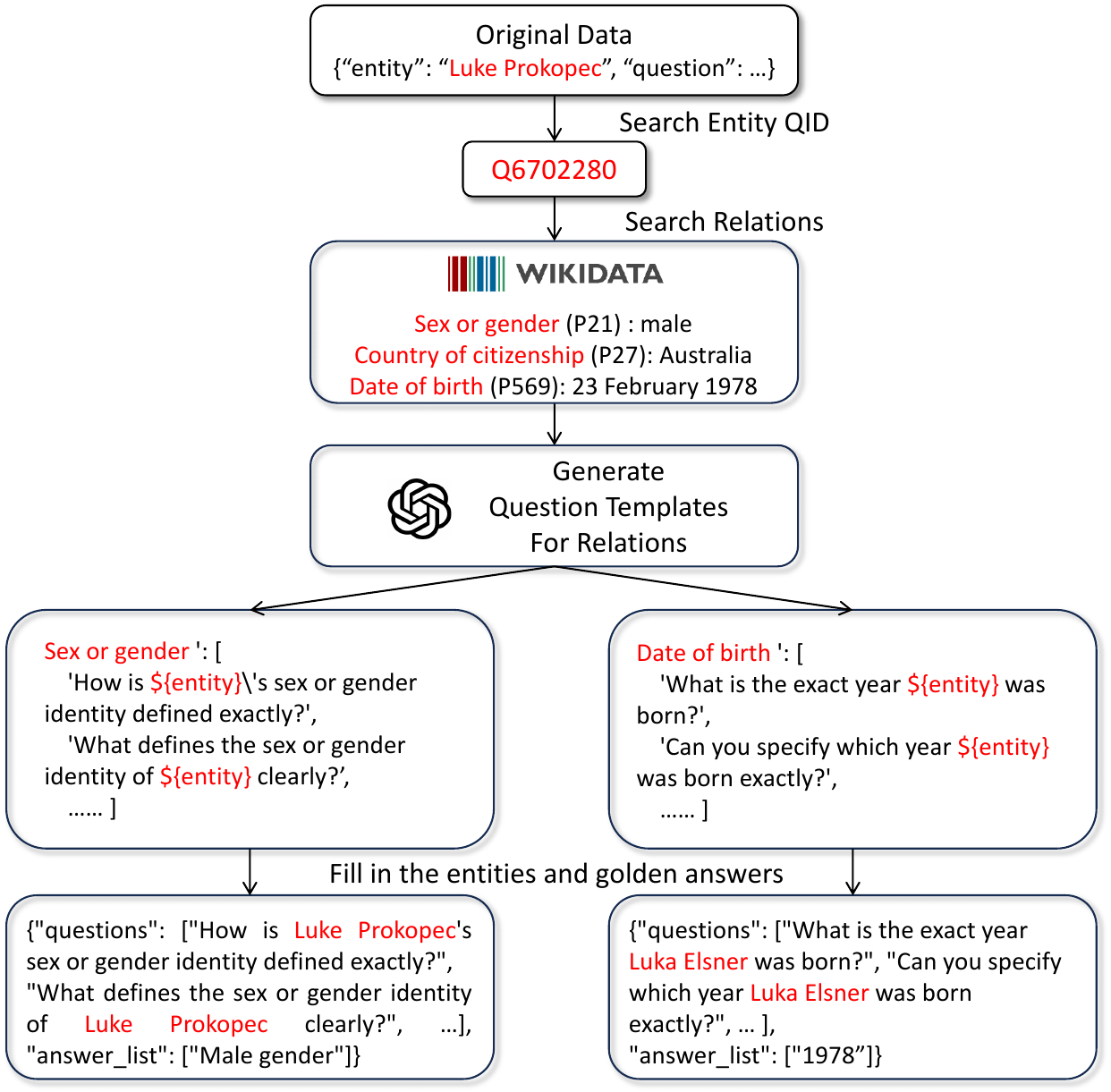}
  \caption{Dataset augmentation workflow. For entity mentions in existing datasets, we retrieve their connected relations and target entities from Wikidata. We then use GPT-4o to generate templates and construct the augmented QA dataset.}
  \label{fig:dataset_augment}
\end{figure}

\textbf{Metrics.}
The primary focus lies in system efficiency and response accuracy. The following metrics are measured:
\begin{itemize}
    \item \textbf{Average Latency (AvgL)}: The average end-to-end retrieval latency in seconds over a batch of queries. It directly represents the system efficiency.
    \item \textbf{Response Accuracy (RA)}: We follow \cite{asai2023} to evaluate answers by checking whether they contain the golden answers. It reflects the retrieval quality.
    \item \textbf{Document Hit Rate}: It is defined as the proportion of queries for which at least one retrieved document contains the golden answer.
    \item \textbf{Response Accuracy conditioned on Draft Acceptance (RA@DA)}: It reports accuracy restricted to accepted drafts, allowing us to isolate and assess the effect of HaS.
    \item \textbf{Draft Acceptance Rate (DAR)}: It indicates how often a homologous query is re-identified and the draft is accepted. 
    \item \textbf{Correct Acceptance Rate (CAR)}: It reflects the precision of the re-identifications and draft acceptances.  
\end{itemize}

\textbf{Baseline methods.}
We compare HaS with three representative categories of state-of-the-art retrieval strategies for acceleration:
\begin{itemize}
    \item We compare HaS with mature ANNS methods, including IVF and ScaNN, under two configurations: (1) using a similar retrieval scope to replace HaS on the edge, and (2) using an optimized scope to replace full-database retrieval on the cloud.
    \item We consider reuse-based methods, including Proximity \cite{bergman2025}, SafeRadius \cite{frieder2024}, and MinCache \cite{haqiq2025}. They skip retrieval by reusing cached results of semantically identical queries.
    \item To illustrate the efficiency of the homology validation, we replace it with the evaluator from CRAG \cite{yan2024}, which calls the LLM to assess each retrieved document.
\end{itemize}

\subsection{Main Results}

\textbf{Validation is crucial to draft quality.}
We first compare HaS with ANNS methods under a similar retrieval scope ratio, with ANNS deployed on the edge as an alternative to HaS. As shown at the top of Table \ref{tab: approximate retrieval}, ANNS methods under a narrow retrieval scope have a high risk of missing critical documents. Without a validation mechanism, they suffer over 10\% accuracy degradation, making their latency savings ineffective. In contrast, HaS can perform validation and fallback to ensure draft quality. Its better response accuracy highlights the critical role of validation in speculative retrieval.

\textbf{ANNS methods and HaS are complementary.}
In practical deployments, ANNS methods are typically configured with a larger, fine-tuned retrieval scope, as a substitute for full-database retrieval on the cloud. While these methods achieve strong performance, they are complementary to HaS. Building on their strengths, HaS can be integrated with them to further improve efficiency. We evaluate this combination at the bottom of Table \ref{tab: approximate retrieval} and observe that HaS brings an additional 7\%–28\% latency reduction on top of their performance.

\begin{table}[htbp]
\centering
\caption{Comparison of HaS and ANNS Methods.
$\spadesuit$: Limited retrieval scope to replace HaS. $\diamondsuit$: Optimized retrieval scope to replace full-database retrieval. $\uparrow$ / $\downarrow$: Higher / lower is better. $\star$: Sampled Queries from the Augmented dataset.}
\begin{tabular}{ccccc}
\hline
\multirow{2}{*}{Methods} & \multicolumn{2}{c}{Granola-EQ $\star$} & \multicolumn{2}{c}{PopQA $\star$} \\
 & AvgL(s) \boldmath$\downarrow$\unboldmath & RA \boldmath$\uparrow$\unboldmath & AvgL(s) \boldmath$\downarrow$\unboldmath & RA \boldmath$\uparrow$\unboldmath \\ \hline
IVF $\spadesuit$ & 0.0921 & 0.4283 & 0.0916 & 0.2569 \\
ScaNN $\spadesuit$ & 0.0802 & 0.4353 & 0.0778 & 0.2524 \\
HaS & 1.0559 & 0.4829 & 0.8725 & 0.2906 \\ \hline
IVF $\diamondsuit$ & 0.5431 & 0.4824 & 0.5432 & 0.2825 \\
\multirow{2}{*}{HaS + IVF $\diamondsuit$} & 0.4603 & 0.4786 & 0.3872 & 0.2784 \\
 & -15.24\% & -0.79\% & -28.73\% & -1.48\% \\ \hline
ScaNN $\diamondsuit$ & 0.3554 & 0.4824 & 0.3553 & 0.2862 \\
\multirow{2}{*}{HaS + ScaNN $\diamondsuit$} & 0.3285 & 0.4790 & 0.2904 & 0.2812 \\
 & -7.55\% & -0.70\% & -18.27\% & -1.76\% \\ \hline
\end{tabular}
\label{tab: approximate retrieval}
\end{table}

\textbf{HaS Leverages Homologous Queries to Achieve Higher Efficiency.}
Existing reuse-based methods attempt to locate a semantically identical query for result reuse. However, such queries are rarely present in a limited cache, limiting their efficiency. As shown in Table \ref{tab:main result}, while maintaining only a small drop in response accuracy, they fail to achieve substantial latency reduction. Fig.\ref{fig:threshold_comparison} further examines their performance across different thresholds. Lowering the threshold relaxes the matching criteria for acceleration, but at the cost of reduced response accuracy. HaS consistently demonstrates Pareto superiority. By exploiting more prevalent homologous queries for validation, HaS achieves greater opportunities for draft acceptance, thus realizing a larger latency reduction.

\begin{table*}[htbp]
\centering
\caption{Performance comparisons. The percentage difference is computed relative to the full-database retrieval. $\text{CRAG}^{\dag}$: Replaces homologous validation with the LLM-based evaluator from CRAG. $\uparrow$ / $\downarrow$: Higher / lower is better. $\star$: Sampled Queries from the Augmented dataset.}
\begin{tabular}{ccccccccccc}
\hline
\multirow{3}{*}{Method} & \multicolumn{5}{c}{Granola-EQ $\star$} & \multicolumn{5}{c}{PopQA $\star$} \\ \cline{2-11} 
 & \multirow{2}{*}{AvgL(s) \boldmath$\downarrow$\unboldmath} & \multirow{2}{*}{\begin{tabular}[c]{@{}c@{}}Doc. Hit \\ Rate \boldmath$\uparrow$\unboldmath\end{tabular}} & \multicolumn{3}{c}{Response Accuracy \boldmath$\uparrow$\unboldmath} & \multirow{2}{*}{AvgL(s) \boldmath$\downarrow$\unboldmath} & \multirow{2}{*}{\begin{tabular}[c]{@{}c@{}}Doc. Hit \\ Rate \boldmath$\uparrow$\unboldmath\end{tabular}} & \multicolumn{3}{c}{Response Accuracy \boldmath$\uparrow$\unboldmath} \\ \cline{4-6} \cline{9-11} 
 &  &  & $\text{Qwen3}_\text{8B}$ & $\text{Llama}_\text{8B}$ & $\text{Mixtral}_\text{7B}$ &  &  & $\text{Qwen3}_\text{8B}$ & $\text{Llama}_\text{8B}$ & $\text{Mixtral}_\text{7B}$ \\ \hline
\begin{tabular}[c]{@{}c@{}}Full-Database\\ Retrieval\end{tabular} & 1.3845 & 0.6457 & 0.4875 & 0.4715 & 0.4806 & 1.3847 & 0.4652 & 0.2970 & 0.2780 & 0.2703 \\ \hline
\multicolumn{11}{c}{\textit{Reuse-based Methods}} \\
\multirow{2}{*}{Proximity} & 1.3186 & 0.6397 & 0.4824 & 0.4656 & \textbf{0.4764} & 1.1328 & 0.4415 & 0.2802 & 0.2614 & 0.2522 \\
 & -4.76\% & -0.92\% & -1.04\% & -1.25\% & \textbf{-0.86\%} & -18.19\% & -5.09\% & -5.66\% & -5.96\% & -6.70\% \\
\multirow{2}{*}{MinCache} & 1.3044 & 0.6329 & 0.4746 & 0.4590 & 0.4679 & 1.0437 & 0.4254 & 0.2676 & 0.2452 & 0.2360 \\
 & -5.78\% & -1.97\% & -2.64\% & -2.65\% & -2.64\% & -24.63\% & -8.55\% & -9.91\% & -11.81\% & -12.68\% \\
\multirow{2}{*}{SafeRadius} & 1.2870 & 0.6355 & 0.4779 & 0.4603 & 0.4718 & 0.9773 & 0.4245 & 0.2649 & 0.2477 & 0.2338 \\
 & -7.05\% & -1.58\% & -1.97\% & -2.38\% & -1.82\% & -29.42\% & -8.74\% & -10.83\% & -10.90\% & -13.51\% \\ \hline
\multicolumn{11}{c}{\textit{LLM for Validation}} \\
\multirow{2}{*}{$\text{CRAG}^{\dag}$} & 1.5196 & 0.6287 & 0.4702 & 0.4549 & 0.4625 & 1.8186 & 0.4557 & 0.2885 & 0.2706 & 0.2542 \\
 & 9.76\% & -2.63\% & -3.54\% & -3.53\% & -3.75\% & 31.33\% & -2.04\% & -2.87\% & -2.65\% & -5.97\% \\ \hline
\multirow{2}{*}{HaS} & \textbf{1.0559} & \textbf{0.6402} & \textbf{0.4829} & \textbf{0.4667} & 0.4755 & \textbf{0.8725} & \textbf{0.4578} & \textbf{0.2906} & \textbf{0.2720} & \textbf{0.2638} \\
 & \textbf{-23.74\%} & \textbf{-0.84\%} & \textbf{-0.94\%} & \textbf{-1.03\%} & -1.06\% & \textbf{-36.99\%} & \textbf{-1.60\%} & \textbf{-2.14\%} & \textbf{-2.14\%} & \textbf{-2.41\%} \\ \hline
\end{tabular}
\label{tab:main result}
\end{table*}

\begin{figure}[htbp]
  \centering
  \includegraphics[width=0.9\linewidth]{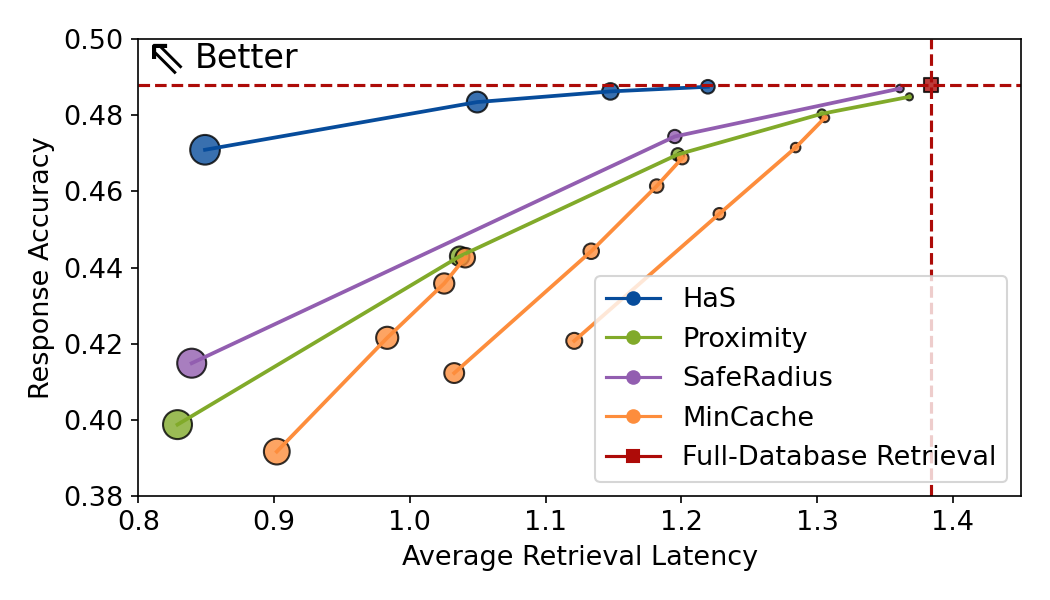}
  \caption{Comparison on varying threshold settings. Point size indicates the percentage of relevant queries found (semantically similar or homologous). MinCache involves two thresholds, resulting in multiple performance curves.}
  \label{fig:threshold_comparison}
\end{figure}

\textbf{Homologous Validation is more efficient.}
To evaluate the benefit of our homologous validation, we replace it with the LLM-based evaluator from CRAG. Some additional metrics are provided in Table \ref{tab: additional performance}. While the LLM-based evaluator more precisely identifies relevant documents, it introduces approximately 0.7 seconds for LLM inference per query. Despite accepting 42.2\% of drafts, the method still results in a 9.76\% increase in average retrieval latency compared to full-database retrieval. Meanwhile, it exhibits weaker confidence when evaluating out-of-distribution data in PopQA. As most drafts are rejected, the latency overhead escalates by 31.33\%. In contrast, by leveraging lightweight validation with minimal overhead, HaS achieves substantially lower average retrieval latency.

\begin{table}[htbp]
\centering
\caption{Comparison on additional performance metrics. L@DA and L@DR denote the per-query retrieval latency (in seconds) upon draft acceptance and rejection.}
\begin{tabular}{c|cccc}
\hline
 & \begin{tabular}[c]{@{}c@{}}DAR \boldmath$\uparrow$\unboldmath\\ (Granola-EQ$\star$)\end{tabular} & \begin{tabular}[c]{@{}c@{}}DAR \boldmath$\uparrow$\unboldmath\\ (PopQA$\star$)\end{tabular} & \begin{tabular}[c]{@{}c@{}}L@DA\\ (s) \boldmath$\downarrow$\unboldmath\end{tabular} & \begin{tabular}[c]{@{}c@{}}L@DR\\ (s) \boldmath$\downarrow$\unboldmath\end{tabular} \\ \hline
$\text{CRAG}^{\dag}$ & 42.20\% & 20.82\% & 0.7006 & 2.1168 \\
HaS & 29.63\% & 43.15\% & 0.0555 & 1.4896 \\ \hline
\end{tabular}
\label{tab: additional performance}
\end{table}

\textbf{HaS is generalizable across datasets.} Table \ref{tab:main result} shows that HaS exhibits steady performance gain across datasets. To further assess its generalizability, we additionally evaluate it on the TriviaQA\footnote{Filtered to contain only well-organized single-hop queries} \cite{triviaqa} and SQuAD \cite{squad} datasets, although these benchmarks deviate from real-world popularity patterns to contain more scattered queries. As shown in Table \ref{tab: generalizable}, such variability is expected to affect all methods that rely on the query popularity. In this setting, by exploiting the prevalence of homologous queries, HaS can still deliver better performance gains. By contrast, reuse-based methods rely on less frequent semantically identical queries, leading to significant performance degradation, particularly on SQuAD.

\begin{table}[htbp]
\centering
\caption{Performance comparison with reuse-based methods on two datasets that deviate from real-world query patterns. \ddag: queries sampled from the original dataset.}
\begin{tabular}{c|cccc}
\hline
\multirow{2}{*}{Method}                                           & \multicolumn{2}{c}{TriviaQA\ddag}         & \multicolumn{2}{c}{SQuAD\ddag}           \\
                                                                  & AvgL(s) \boldmath$\downarrow$\unboldmath              & RA \boldmath$\uparrow$\unboldmath               & AvgL(s) \boldmath$\downarrow$\unboldmath             & RA \boldmath$\uparrow$\unboldmath               \\ \hline
\begin{tabular}[c]{@{}c@{}}Full-Database\\ Retrieval\end{tabular} & 1.3843            & 0.7615           & 1.3846           & 0.2787           \\
\multirow{2}{*}{Proximity}                                        & 1.3460            & 0.7601           & 1.4135           & 0.2779           \\
                                                                  & -2.77\%           & -0.19\%          & 2.09\%           & -0.28\%          \\
\multirow{2}{*}{MinCache}                                         & 1.3438            & 0.7574           & 1.4204           & 0.2781           \\
                                                                  & -2.93\%           & -0.54\%          & 2.59\%           & -0.21\%          \\
\multirow{2}{*}{Safe Radius}                                      & 1.3296            & 0.7572           & 1.4918           & 0.2758           \\
                                                                  & -3.96\%           & -0.56\%          & 7.75\%           & -1.04\%          \\
\multirow{2}{*}{Ours}                                             & \textbf{1.0810}   & \textbf{0.7603}  & \textbf{1.2910}  & \textbf{0.2782}  \\
                                                                  & \textbf{-21.91\%} & \textbf{-0.15\%} & \textbf{-6.76\%} & \textbf{-0.19\%} \\ \hline
\end{tabular}
\label{tab: generalizable}
\end{table}

\subsection{Analysis on the Fuzzy Channel}

\textbf{Fuzzy Channel Enhances Validation Precision and Draft Quality.}
The fuzzy channel in HaS serves a dual role: enhancing homologous validation and improving draft quality. We conduct an ablation study under two settings: (i) removing it from validation, where only documents from the cache channel are used for homologous verification\footnote{To prevent cold-start failures due to repeated matches, only in this experiment, we pre-fill the cache with random queries.}; and (ii) disabling it for draft enhancement, where accepted drafts exclude documents from the fuzzy channel. Results are illustrated in Table \ref{tab: ab1}. Excluding the fuzzy channel during validation abnormally increases DAR but reduces CAR. It indicates that the system mistakenly matches non-homologous queries, thus undermining validation reliability and reducing response accuracy. When removing the fuzzy channel for draft enhancement, RA@DA drops from 0.4907 to 0.4504. The degradation is more pronounced under unreliable validation settings. It underscores its importance in enhancing draft quality, particularly when documents from the cache channel are unreliable.

\begin{table}[htbp]
\centering
\caption{Impact of incorporating the fuzzy channel for validation (V) and draft enhancement (E).}
\begin{tabular}{cc|ccccc}
\hline
V & E & AvgL(s) \boldmath$\downarrow$\unboldmath & RA \boldmath$\uparrow$\unboldmath & DAR \boldmath$\uparrow$\unboldmath & CAR  \boldmath$\uparrow$\unboldmath & RA@DA  \boldmath$\uparrow$\unboldmath \\ \hline
$\times$  & $\times$  & \multirow{2}{*}{0.1411} & 0.2308 & \multirow{2}{*}{92.80\%} & \multirow{2}{*}{34.77\%} & 0.2076 \\
$\times$ & \checkmark &  & 0.4294 &  &  & 0.4216 \\
\checkmark & $\times$ & \multirow{2}{*}{0.9926} & 0.4673 & \multirow{2}{*}{34.67\%} & \multirow{2}{*}{88.77\%} & 0.4504 \\
\checkmark & \checkmark &  & 0.4813 &  &  & 0.4907 \\ \hline
\end{tabular}
\label{tab: ab1}
\end{table}

\textbf{Compressing the Fuzzy Channel Enables Resource-Constrained Deployment.}
Although the fuzzy channel performs ANNS to retrieve only a small scope of the database, it still has to load the entire database, incurring considerable storage overhead. To enable deployment on resource-constrained edge devices, HaS can compress the fuzzy channel by loading only a smaller subset of the database. We evaluate the system performance under different subset proportions to examine this characteristic. On top of Table \ref{tab: ab2}, with a fixed threshold $\tau$, the performance shows a decline as the proportion shrinks. Since it becomes less likely to include documents relevant to the current query, its validation and enhancement ability are weakened. However, this degradation can be mitigated by adjusting the threshold $\tau$. The lower half of the table shows the fine-tuned performances. Even with the proportion reduced to just 1\% of the full database, tuning the threshold to $\tau = 0.6$ enables HaS to maintain comparable performance, incurring only a 4.4\% increase in AvgL and a 2.8\% drop in RA. These findings confirm the robustness of HaS under aggressive compression, enabling practical and efficient deployment on lightweight devices.

\begin{table}[htbp]
\centering
\caption{Impact of the proportion of the full database used for fuzzy channel construction and matching threshold $\tau$ on system performance. Fine-tuning $\tau$ enables maintaining comparable performance as the fuzzy channel compressed.}
\begin{tabular}{cc|cccc}
\hline
\% & $\tau$ & AvgL(s) \boldmath$\downarrow$\unboldmath & RA \boldmath$\uparrow$\unboldmath & DAR \boldmath$\uparrow$\unboldmath & RA@DA  \boldmath$\uparrow$\unboldmath \\ \hline
1 & \multirow{4}{*}{0.2} & 0.4982 & 0.3236 & 67.38\% & 0.2361 \\
10 &  & 0.8223 & 0.4342 & 45.07\% & 0.3643 \\
50 &  & 0.9736 & 0.4696 & 34.73\% & 0.4414 \\
100 &  & 1.0559 & 0.4829 & 29.63\% & 0.4892 \\ \hline
1 & 0.6 & 1.0964 & 0.4698 & 25.71\% & 0.4323 \\
10 & 0.4 & 1.0573 & 0.4706 & 28.70\% & 0.4440 \\
50 & 0.3 & 1.0776 & 0.4790 & 27.49\% & 0.4732 \\
100 & 0.2 & 1.0559 & 0.4829 & 29.63\% & 0.4892 \\ \hline
\end{tabular}
\label{tab: ab2}
\end{table}

\subsection{Parameter Analysis}

\textbf{Varying $k$ Reveals a U-Shaped Performance Impact.}
The number of retrieved documents $k$ affects both homologous validation and response generation. Since its U-shaped influence on generation has been discussed in prior work \cite{jin2024a}, we focus on its role in homologous validation. Fig.\ref{fig:k_selections} shows that, unlike random queries, homologous queries tend to share retrieved documents at top rank positions. A smaller $k$ can better capture these overlaps and yield higher validation precision. As evidenced by the left side of Fig.\ref{fig:k_comparison_combined}, decreasing $k$ filters noise on lower rank positions, thus bringing better re-identification precision and higher CAR. Combining both aspects, the influence of $k$ in HaS follows a U-shaped trend, as demonstrated by the right side of Fig.\ref{fig:k_comparison_combined}. While a smaller $k$ improves validation accuracy, it risks omitting factual documents. $k=10$ strikes the best balance, whereas larger $k$ slightly degrades performances due to increased noise and weaker validation.

\begin{figure}[htbp]
  \centering
  \includegraphics[width=0.9\linewidth]{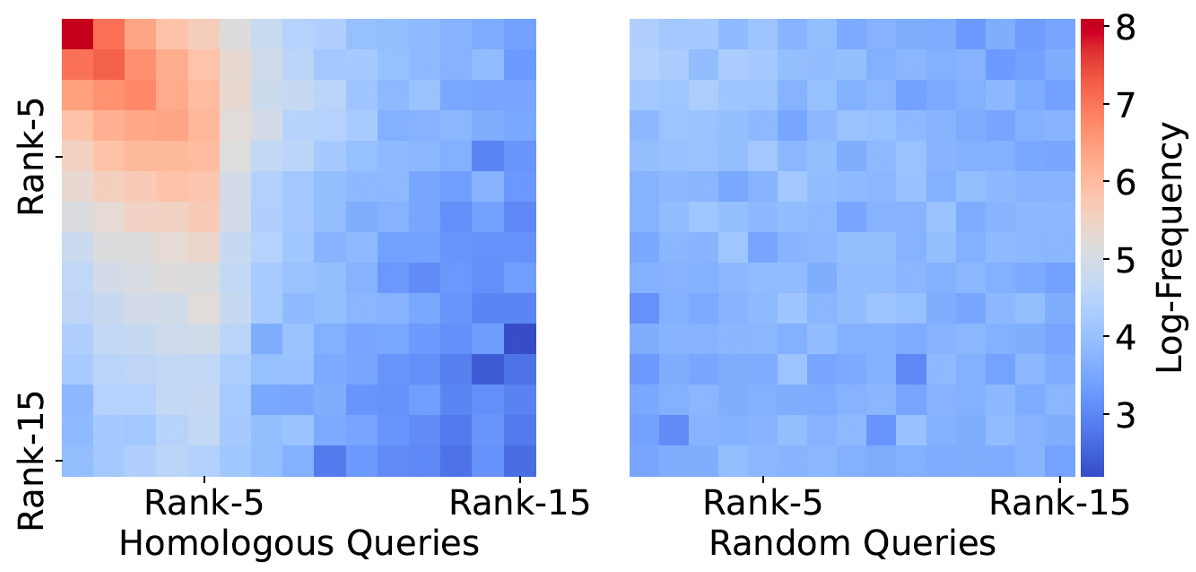}
  \caption{Joint distribution of document rankings for two homologous queries or two random queries. Warm-colored pixels at the top-left corner indicate that the two queries have many retrieved documents in common.}
  \label{fig:k_selections}
\end{figure}

\begin{figure}[htbp]
  \centering
  \includegraphics[width=0.95\linewidth]{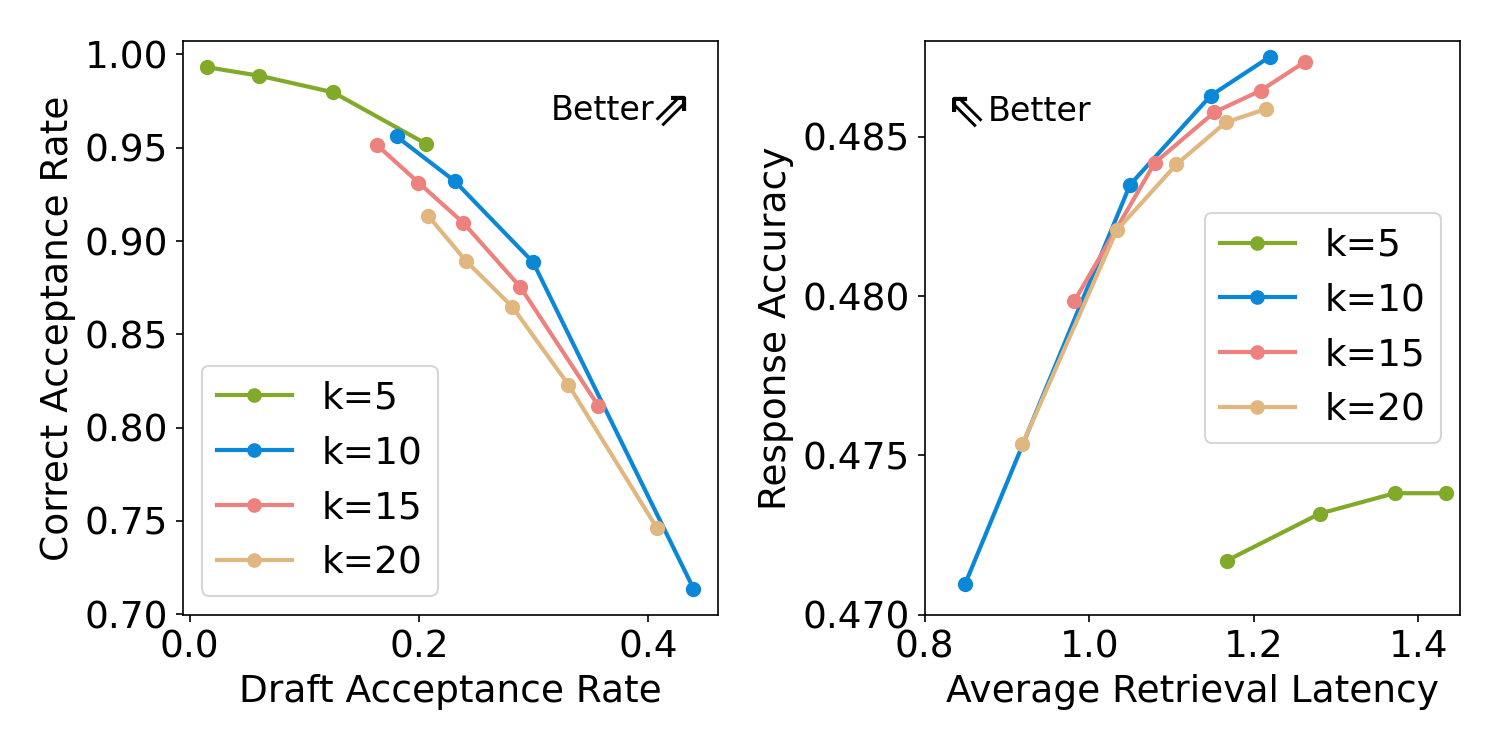}
  \caption{System performance under different $k$ and varying thresholds.}
  \label{fig:k_comparison_combined}
\end{figure}

\textbf{Increasing $\tau$ trades latency for accuracy.}
The threshold $\tau$ distinguishes homologous queries from others during the re-identification process. Increasing $\tau$ enforces stricter homology validation, directly affecting the latency–accuracy trade-off. As shown in Table \ref{tab: diff encoder and tau}, higher $\tau$ values generally increase average latency while improving response accuracy, as a stricter condition is enforced to re-identify homologous queries and accept drafts. However, excessively high $\tau$ suffers from diminishing returns, causing unnecessary latency for marginal accuracy gains. In our setup, $\tau = 0.2$ achieves a satisfactory balance.

\textbf{HaS is robust to the choice of encoders.} 
In retrieval, the underlying encoder serves to extract features for ENNS, and variations across encoders can result in differing retrieval quality and preferences. In addition to the classical \texttt{Contriever} \cite{izacard2021contriever}, we select two recent encoders, \texttt{BGE-Large-en-v1.5} \cite{bge_embedding} and \texttt{e5-Base-v2} \cite{wang2022text}. Table \ref{tab: diff encoder and tau} demonstrates that HaS maintains robust performance across these encoders. It consistently reduces average latency compared to full-database retrieval while preserving comparable response accuracy. Moreover, under the same $\tau$, HaS exhibits similar trade-offs across different encoders. This further highlights its plug-and-play nature. Switching encoders does not require complex parameter tuning.

\begin{table}[htbp]
\centering
\caption{Impact of the encoder and threshold $\tau$ on system performance. HaS is robust to different encoders. Higher values of $\tau$ trade off average latency for improved response accuracy.}
\begin{tabular}{ccc|cc}
\hline
 &  & $\tau$ & AvgL(s) \boldmath$\downarrow$\unboldmath & RA \boldmath$\uparrow$\unboldmath \\ \hline
\multirow{4}{*}{Contriever} & \begin{tabular}[c]{@{}c@{}}Full-Database\\ Retrieval\end{tabular} & - & 1.3845 & 0.4875 \\
 & \multirow{3}{*}{HaS} & 0.1 & 0.8491 & 0.4709 \\
 &  & 0.2 & 1.0496 & 0.4835 \\
 &  & 0.3 & 1.1477 & 0.4863 \\ \hline
\multirow{4}{*}{BGE-Large} & \begin{tabular}[c]{@{}c@{}}Full-Database\\ Retrieval\end{tabular} & - & 1.4191 & 0.4971 \\
 & \multirow{3}{*}{HaS} & 0.1 & 0.9076 & 0.4788 \\
 &  & 0.2 & 1.1528 & 0.4922 \\
 &  & 0.3 & 1.2708 & 0.4954 \\ \hline
\multirow{4}{*}{e5-Base} & \begin{tabular}[c]{@{}c@{}}Full-Database\\ Retrieval\end{tabular} & - & 1.4923 & 0.5003 \\
 & \multirow{3}{*}{HaS} & 0.1 & 0.9256 & 0.4844 \\
 &  & 0.2 & 1.0586 & 0.4932 \\
 &  & 0.3 & 1.2425 & 0.4965 \\ \hline
\end{tabular}
\label{tab: diff encoder and tau}
\end{table}

\textbf{Larger cache size leads to better system efficiency.}
To isolate the impact of cache size on efficiency, we adjust thresholds to maintain comparable response accuracy across settings. As shown in Table \ref{tab: cache size}, enlarging the cache consistently enhances efficiency by increasing the likelihood of re-identifying homologous queries and draft acceptance. Meanwhile, the overhead introduced by a larger cache is minimal. For queries with draft acceptance, retrieval latency remains nearly constant. When the draft is rejected, FIFO-based updates incur only an additional $\sim$12 ms per query, which is negligible compared to the latency saved. Furthermore, the memory footprint increases linearly at a modest rate, approximately 5 MB for every additional 1,000 entries, which remains well within the capacity of modern server infrastructures. These results highlight that enlarging the cache enhances system effectiveness without incurring significant resource overhead.

\begin{table}[htbp]
\centering
\caption{System efficiency under varying cache size $H_{\text{max}}$. L@DA and L@DR denote the per-query retrieval latency (in seconds) upon draft acceptance and rejection. Mem represents the memory footprint (in MB) of the cache.}
\begin{tabular}{cccccc}
\hline
$H_{\text{max}}$ & AvgL(s)\boldmath$\downarrow$\unboldmath & DAR\boldmath$\uparrow$\unboldmath & L@DA(s)\boldmath$\downarrow$\unboldmath & L@DR(s)\boldmath$\downarrow$\unboldmath & Mem(MB) \\ \hline
2000 & 1.2702 & 14.30\% & 0.0547 & 1.4778 & 15.10 \\
3000 & 1.1876 & 20.24\% & 0.0549 & 1.4828 & 20.45 \\
4000 & 1.1165 & 25.31\% & 0.0552 & 1.4892 & 26.59 \\
5000 & 1.0559 & 29.63\% & 0.0555 & 1.4896 & 31.38 \\
\hline
\end{tabular}
\label{tab: cache size}
\end{table}

\subsection{Use Case}
\label{use case}
\textbf{I. Case Study of HaS on Simple Queries.}
We present an illustrative case study in Fig.\ref{fig:case_study}. For a query, a two-channel fast retrieval is first performed, followed by the homology validation. Due to the entity-alignment preference in retrieval, although the retrieval results of the cached query focus on the occupation attribute, some of these documents (Documents 2 and 3) are also included in the draft. This correctly confirms their homology and leads to the acceptance of the draft. This facilitates the efficiency and accuracy of our validation mechanism.

This example also highlights the advantages of the fuzzy channel. First, the fuzzy channel enhances validation reliability. As illustrated in Fig.\ref{fig:case_study_1}, when only 2 documents are relevant, rank-based retrieval on the cache channel returns a noisy Document 4 among the Top-3 results. Without the fuzzy channel, this document would be included in the draft, erroneously inflating the homology score of the linked irrelevant query, thus increasing the risk of incorrect re-identification. Incorporating the fuzzy channel provides more relevant documents, such as Document 1, to replace noisy ones for better validation reliability. Second, the fuzzy channel complements the cache channel by contributing additional valuable documents. In Fig.\ref{fig:case_study_2}, assuming the cache channel lacks the informative Document 2, the fuzzy channel contributes the supplemental Document 1, safeguarding the system against accuracy degradation.

\begin{figure}[htbp]
\centering
\subfloat[Two-channel fast retrieval.\label{fig:case_study_1}]{
\includegraphics[width=0.9\linewidth]{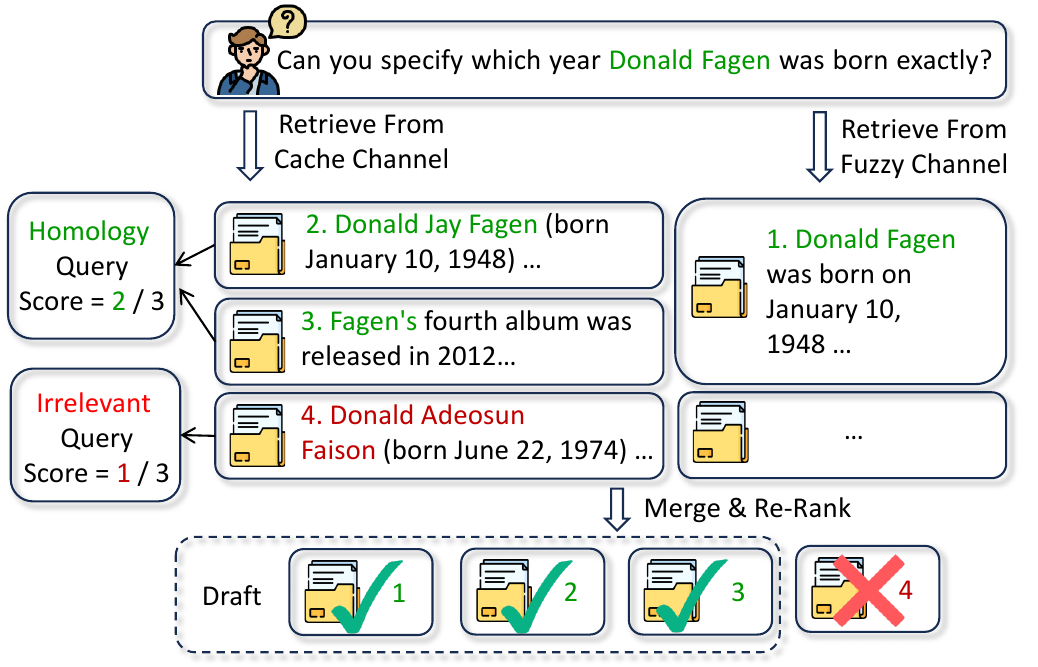}
}
\hfill
\subfloat[Homology validation.\label{fig:case_study_2}]{
\includegraphics[width=0.9\linewidth]{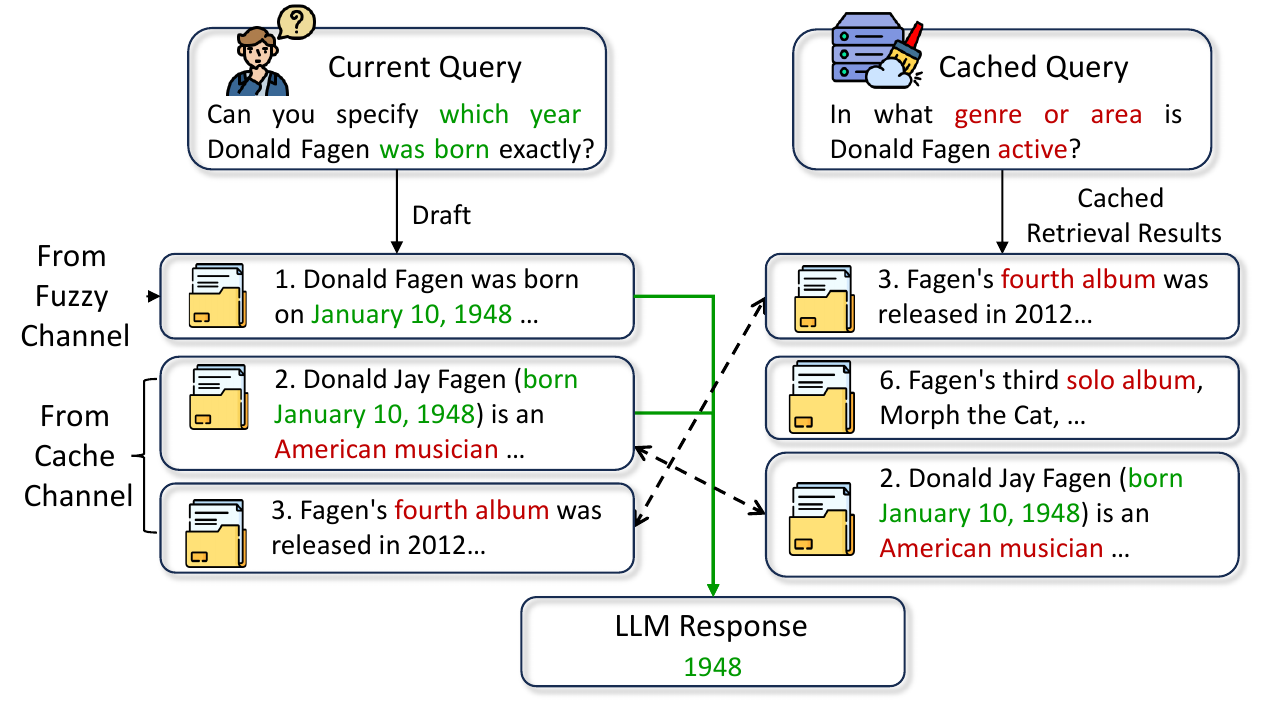}
}
\caption{An illustrative example for a case study.}
\label{fig:case_study}
\end{figure}

\textbf{II. Plugging HaS into modern RAG pipelines to solve complex queries.} 
In previous sections, we evaluated HaS within a basic RAG pipeline for simple queries. This does not mean that HaS is restrictive. As a pluggable component, HaS can be seamlessly integrated into any RAG pipelines without much effort. Practically, one only needs to direct queries to our HaS for retrieval. No modification to the pipeline implementation is required. With built-in query decomposition mechanisms in modern pipelines, complex queries involving multiple entities and attributes can be solved iteratively.

\begin{figure}[htbp]
  \centering
  \includegraphics[width=0.9\linewidth]{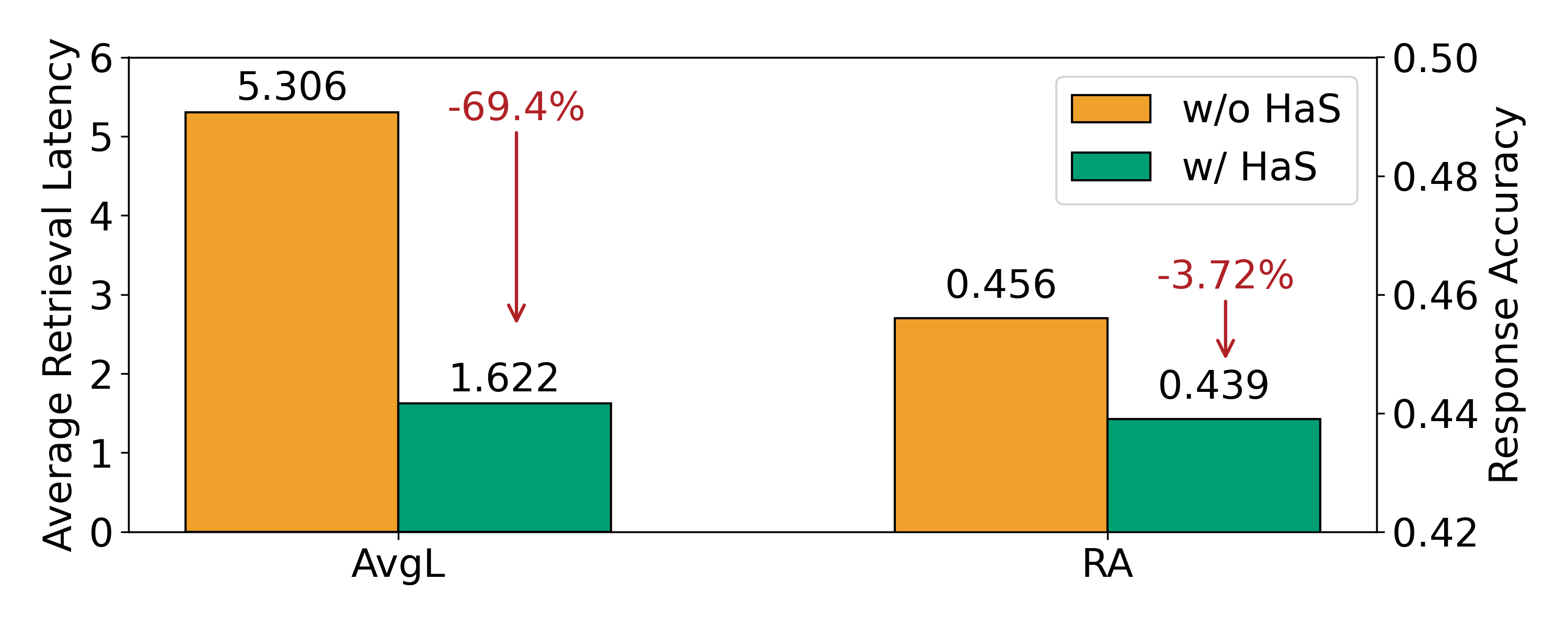}
  \caption{System performance with and without HaS in the Auto-RAG pipeline on complex queries.}
  \label{fig:case study}
\end{figure}

As a case study, we integrate HaS into Auto-RAG \cite{yu2024b}, a Chain-of-Thought-based RAG pipeline. It performs query decomposition and iterative retrieval to handle complex queries. We similarly constructed a 2-hop dataset for evaluation. Fig.\ref{fig:case study} shows that HaS reduces average retrieval latency by 69.4\% with only 3.72\% accuracy degradation. The high frequency of homologous patterns among decomposed sub-queries boosts the draft acceptance rate, bringing significant acceleration. Fig.\ref{fig:case study multi-hop} presents an example. For a complex query, Auto-RAG decomposes the query into sub-queries through reasoning. For the first sub-query, since a homologous query is re-identified, the draft is accepted for acceleration. Auto-RAG then proceeds to the next hop. At this point, the draft is rejected, and the system falls back to the full-database retrieval. These results demonstrate that HaS functions as a practical drop-in solution for accelerating modern RAG systems.

\begin{figure}[htbp]
  \centering
  \includegraphics[width=0.8\linewidth]{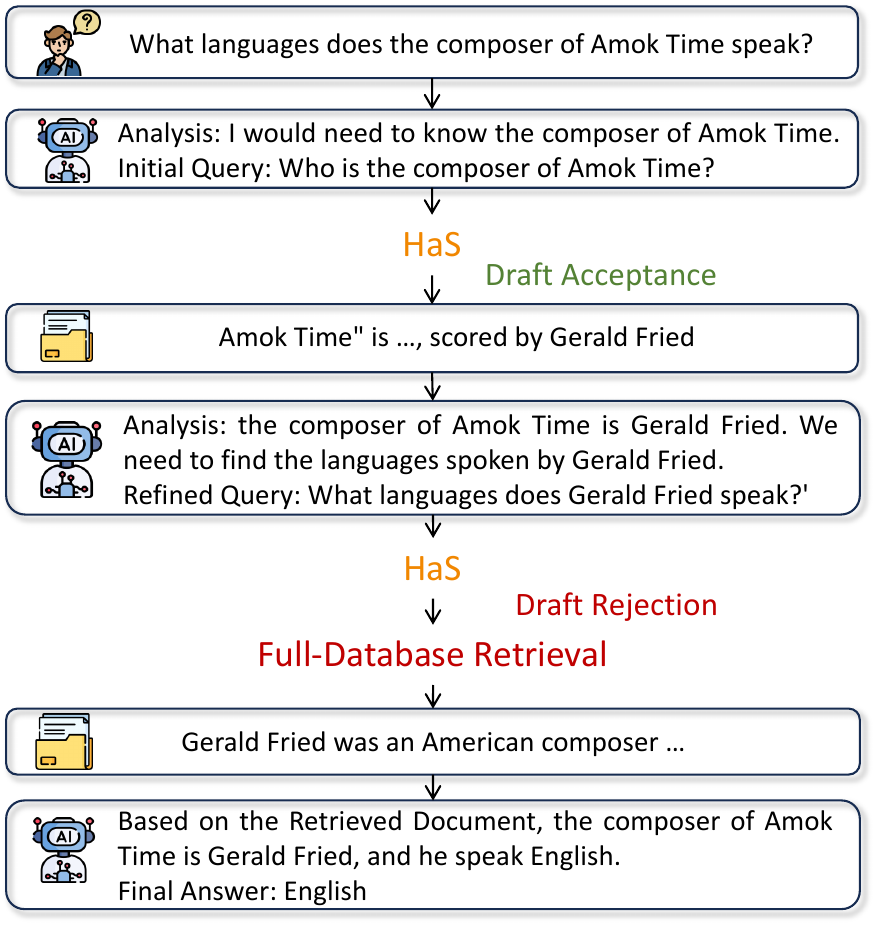}
    \caption{An illustrative case study demonstrating how HaS can be plugged into Auto-RAG to solve multi-hop queries.}
  \label{fig:case study multi-hop}
\end{figure}

\section{Related Works}
\subsection{Empowering LLMs with an Efficient RAG System}
RAG-empowered LLM systems have attracted extensive research effort, particularly on efficiency. A primary direction focuses on reducing prefilling latency caused by extra context length. At the lexical level, methods like RECOMP \cite{xu2023e} try to filter out irrelevant content to shorten the token length of prompts. In \cite{wang2024i}, documents are distributed across multiple specialists for parallel accelerated generation, followed by a generalist who selects the best result. At the embedding level, works like xRAG \cite{cheng2024b} compress features of multiple tokens into a single one as input to the downstream LLM. Several studies explore caching key-value (KV) features to bypass prefilling. TurboRAG \cite{lu2024} precomputes KV features of documents, and directly concatenates retrieved ones to skip prefill. However, the omission of cross-document attention risks diminishing LLM response accuracy. Considering this limitation, RAGCache \cite{jin2024} employs a prefix tree structure for KV caching, matching only the longest prefix. CacheBlend \cite{yao2024a} selectively recomputes part of tokens exhibiting high KV deviation, thus mitigating accuracy degradation with minimal computational costs.

\subsection{Accelerating Retrieval over Large-Scale Corpora}
Retrieval latency in large-scale RAG systems has become a bottleneck \cite{quinn2025}. Industrial solutions like Faiss typically employ ANNS methods such as IVF, HNSW \cite{hnsw}, and ScaNN \cite{scann}, which aim to retrieve within a smaller scope for acceleration. Despite their efficiency, they inevitably sacrifice accuracy. To mitigate this issue, some studies introduce evaluators to assess retrieved documents \cite{hei2024, yan2024, zhu2024}, though the associated inference latency diminishes the overall efficiency gains. Federated retrieval adheres to a similar underlying concept. They partition the large corpus into shards. A router selects relevant shards for parallel retrieval \cite{li2024c, shojaee2025}. Although it brings retrieval acceleration, the computational cost associated with the router remains substantial.

Another intuitive optimization is results reuse. A caching mechanism stores historical queries and their corresponding results, allowing reuse when sufficient semantic similarity with current queries is established. Common approaches quantify semantic similarity through cosine similarity between query embeddings \cite{bergman2025}. A safe radius criterion has been proposed in \cite{frieder2024}, defining a hyperball covering the cached query and retrieved results, with reuse permitted if the current query lies within this region. A three-tier matching framework is introduced in \cite{haqiq2025}, featuring a resemblance matching step that assesses similarity using lexical-level Jaccard similarity over MinHash signatures. However, in real-world scenarios, users tend to issue queries about different attributes of the same entity. This diversity conflicts with the strict semantic similarity requirement, limiting acceleration gains.

\section{Conclusion}

We present HaS, a speculative retrieval-based framework that leverages historical homologous queries for lightweight draft validation. HaS first performs two-channel fast retrieval from a cache channel and a fuzzy channel. They jointly define a much narrower retrieval range than the full database to provide draft documents with much lower latency. By trying to re-identify a homologous query from the cache, HaS determines the draft’s relevance and skips full retrieval if validated. Experiments show that HaS significantly reduces average retrieval latency with minimal accuracy degradation, consistently outperforming other methods. As a modular retriever, HaS can also be seamlessly integrated into existing RAG pipelines, delivering further efficiency gains as modern pipelines continue to grow in depth and complexity.

\clearpage

\section*{AI-Generated Content Acknowledgement}
Portions of the text in this paper are refined using OpenAI GPT-5 and Google Gemini-2.5-Flash to improve clarity, grammar, and readability. The AI systems are used solely for language polishing. All ideas, figures, analyses, results, and conclusions are entirely those of the authors.

\bibliographystyle{IEEEtran}  
\bibliography{ref}

\end{document}